\documentclass{article}

\usepackage{PRIMEarxiv}

\usepackage[utf8]{inputenc} % allow utf-8 input
\usepackage[T1]{fontenc}    % use 8-bit T1 fonts
\usepackage{hyperref}       % hyperlinks
\usepackage{url}            % simple URL typesetting
\usepackage{booktabs}       % professional-quality tables
\usepackage{amsfonts}       % blackboard math symbols
\usepackage{nicefrac}       % compact symbols for 1/2, etc.
\usepackage{microtype}      % microtypography
\usepackage{lipsum}
\usepackage{fancyhdr}       % header
\usepackage{graphicx}
\usepackage{caption}
\usepackage{subcaption}
\graphicspath{{media/}}     % organize your images and other figures under media/ folder

\title{A Survey on SDN \& SDCN Traffic Measurement: Existing Approaches and Research Challenge
}

\author{
  MD Samiul Islam\\
  University of Alberta \\
  Edmonton, Canada\\
  \texttt{mdsamiul@ualberta.ca} \\
   \And
  Mojammel Hossain \\
  North South University \\
  Dhaka, Bangladesh\\
  \texttt{mojammel.hossain@northsouth.edu} \\
   \And
  Mohammed AlMukhtar \\
  University of Baghdad \\
  Baghdad, Iraq\\
  \texttt{mohammed.abdul@cc.uobaghdad.edu.iq} \\
}

\begin{document}
\maketitle

\begin{abstract}
Software Defined Network (SDN) is the next generation network that decouples the control plane from the data plane of forwarding devices by utilizing the OpenFlow protocol as a communication link between the data plane and the control plane. However, there are some security issues might be in actions on SDN that the attackers can take control over the SDN control plane. Thus, traffic measurement is a fundamental technique of protecting SDN against the high-security threats such as DDoS, heavy hitter, superspreader as well as live video calling, QoS control, high bandwidth requirement, resource management are also inevitable in SDN/Software Defined Cellular Network (SDCN). In such a scenario, we survey SDN traffic measurement solutions, in order to assess how these solutions can make a secured, efficient and robust SDN/SDCN architecture. In this paper, various types of SDN traffic measurement solutions have been categorized based on network applications behaviour. Furthermore, we find out the challenges related to SDN/SDCN traffic measurement and future scope of research, which will guide to design and develop more advanced traffic measurement solutions for a scalable, heterogeneous, hierarchical and widely deployed SDN/SDCN in future prospects. More in details, we list out kinds of practical machine learning (ML) approaches to analyze how we can make improvement in the traffic measurement performances. We conclude that using ML in SDN traffic measurement solutions will give benefit to get secured SDN/SDCN network in complementary ways.
\end{abstract}

% keywords can be removed

\section{Introduction}

To solve the above problem, OpenFlow is proposed as an online communication protocol which is able to control the forwarding plane of an online switch or router. Also,  OpenFlow API and the stateful data plane\cite{sun2017sdpa} are very widespread among the researchers as they are straightforward to collaborate with software controller. Introductory vendors that upheld Open-Flow included HP, NEC, some others, and this rundown has since extended significantly. The combination of SDN \cite{SoftwareDefinedNetworks:TheNewNormofNetworks.} and OpenFlow \cite{opendaylight} is currently the simplest way to perform Quality of Service (QoS) estimation, whenever and wherever using self-guided, self-tuning components that persistently screen and measure network execution and respond quickly to issues \cite{majeed2021spike_UOB2}. With the program based OpenFlow API, a wide range of controllers stages has been developed \cite{Erickson.} for programmers to make numerous applications such as dynamic access control \cite{Casado2007}, network virtualization \cite{blenk_network_virtualization}, energy efficient networking \cite{Heller2010}, consistent virtual-machine relocation and client versatility.

In traditional network, devices such as switches, routers are inflexible and cannot deal with different types of network traffic due to the underlying hardwired implementation of routing rules and other obstacles. This programmable network can be designed for the need of the network operator, which may deploy its own rule of bandwidth, latency, packet missing and inactivity to help the decent variety necessities of cutting-edge network applications and administrations \cite{T}. Then, network administrators take advantage of SDN scalability to offer dynamic QoS that ensures benefit quality between endpoints, paying little concern on the way conveys that administration. Because the traditional networking devices are closed without programmability, each vendor is developing the code for the network protocol. If there is an error in their code and the fault occurred due to this error, then the network administrator/developer is unable to unfold such a fault, which makes SDN unsecured and unreliable to the clients. Although the hybrid deployment of SDN\cite{amin2018hybrid} has many advantages including adaptability to budget constraints, central programmability of the network, it is limited by its structure. Thus, we need to find out the optimal solutions to make the SDN more powerful with a lot of network functionalists.

Therefore, SDN traffic measurement \cite{akyildiz2016research} comes into consideration to break down the complexity of SDN unresolved problems, which also supports different types of tasks to OpenFlow controllers to utilize programmable interfaces. These software defined measurement solutions help to support the diversity requirements of next generation network applications and services by providing consistent traffic measurement of flow parameters such as bandwidth, packet loss, and latency. Also, whether the path carries that service, the flexibility of software defined measurement still gives the network operator the capability to offer dynamic QoS that guarantees service quality between endpoints. Moreover, an indirect, non-intrusive, and statistical way, which in some cases cannot be measured in traditional large networks, are allowed in SDN traffic measurement to infer several characteristics. Although it is quite a mature technology in these days, it is still challenging for programmers to analyze or predict the traffic measurement in future to provide dynamic QoS to guarantees service quality.

However, with the development of machine learning (ML) algorithms, we can make decisions on plenty of tasks using previous knowledge. In general, statistical machine learning, deep learning (DL) and deep reinforcement learning (DRL) are the main methods to deal with the challenge of future prediction. Here, ML is a widely used way to measure the traffic data and it can give the network administrator of future recommendation. Moreover, DL is a recent method which performs more efficient and accurate than the classic ML approaches and can crack out more represented features to make a prediction. Besides, different from these two approaches, DRL concerns more on how to lead the agents to take action for achieving the final goals, and it will interact with the environment to make decisions. With the analysis of the traffic data, SDN traffic measurement among the network can be predicted and then programmers can prevent attackers to control the controller of the network, which provides satisfying service and guarantees the QoS.

In this paper, we present a survey of the research relating to solutions in SDN traffic measurement that have been carried out until 2018. We first describe the current issues about the SDN, SDCN and traffic measurement, and then analyze the existing approaches about their aspects and contributions. To further improve the performance on the SDN traffic measurement, we summarize out the future scopes in ML-based optimization and networking deployment. The main contribution of this paper is as follows:

\begin{itemize}
    
    \item   We find out the basic requirements to design an SDN traffic measurement framework, which can be used as a standard design method in this area.
    \item   We provide a brief literature review on the SDN \& SDCN measurement and categorize them by the network applications and learning behaviour, then we also summarize the future scopes of research with the existing challenges.
    \item   To further improve the measurement of traffic data, we summarize the probability of how to apply machine learning, deep learning or deep reinforcement learning to the traffic measurement. More in details, we analyze previous machine learning-based methods and then transfer the basic idea into the other applications with different approaches.
    \item   At last, we conclude that machine learning is the best choice to design a framework, where we can use a set of algorithms as a measurement library for SDN traffic measurement of the next generation heterogeneous, complex, hierarchical network.
    
\end{itemize}

The remainder of this paper is organized as follows. In Section II, we first give a short brief description of SDN and OpenFlow about its development and some measurement issues and then describe the currently existed cellular network in Section III and discuss how to solve the security issues to improve the quality of SDCN. After this, in Section IV, we introduce the SDN traffic measurement in details. In addition, we categorize the previous state-of-the-art traffic measurement solutions on SDN/SDCN into 4 different segments based on the network applications behaviour in Section V. And we also categorize a few ML SDN traffic measurement solutions based on the learning behaviour to show the advantages of using ML in Section VI. After that, we list out future prospects of ML-based networking and 9-future research directions opened in SDN/SDCN traffic measurement in Section VII and Section VIII respectively. At last, Section IX summarizes the conclusions.

\section{Software Defined Network}

This section represents basic description of SDN, including a momentary introduction and the proposed architectures of SDN. In addition, we describe the widely accepted southbound interface of OpenFlow protocol and the corresponding working flow in SDN environment.

\subsection{Baseline of the SDN Network}

In short, SDN is a network, which separates the control logic from data plane to concentrate into a new control plane, that all command is logically centralized in a controller and it can logically control all data plane equipment. Among the data exchanging, flow is a concept of a sequence of packets, which are sent to the destination from the source. 
% This packet forwarding by data plane devices is flow-based rather than destination based, which functionality from data plane is expelled from network devices that will wind up straightforward packet forwarding components. 
The pattern matching (which can have only destination address) is used for flow rule installation at the switches, and the switches forward the data packets based on pattern matching. 
Also, the arrangement of packet field extensively defines a flow esteems going about as a matching model and an arrangement of activities, and all packets of a flow get indistinguishable administration approaches at the forwarding devices \cite{Newman.1998,N.Gudeetal..2008}. 

On the other hand, the flow reflection permits bringing together the conduct of various kinds of network devices, including routers, switches, firewalls, middle-boxes \cite{Jamjoom.}, and flow programming empowers remarkable adaptability, constrained just to the capacities of the actualized flow tables \cite{McKeown.2008}. Beside, control plane is moved to an outside substance as the SDN controller or NOS. 
% The NOS is a software stage, motivation of which is in this way like of a conventional working framework that keeps running on control server and gives the critical assets and deliberations to encourage the programming of forwarding devices in light of a logically incorporated. 
The NOS is a software stage, which is aimed to keep running on control server and to give the critical assets and deliberations to encourage the programming of forwarding devices in light of a logically incorporated.
This network is programmable through software applications running over the NOS that interfaces with the primary data plane devices, as the traditional network depends on the data plane, where all logic are associated with switches/routers. Also, a network administrator needs to program and monitor every data plane device one after another, which is time-consuming and lengthy. Fig.~\ref{simple_sdn_architecture}, shows the simple SDN architecture of how control and data plane are separated. There are some terminologies related to SDN and SDN traffic measurement as following:

\begin{figure}[t!]
\centering
\includegraphics[width=12cm]{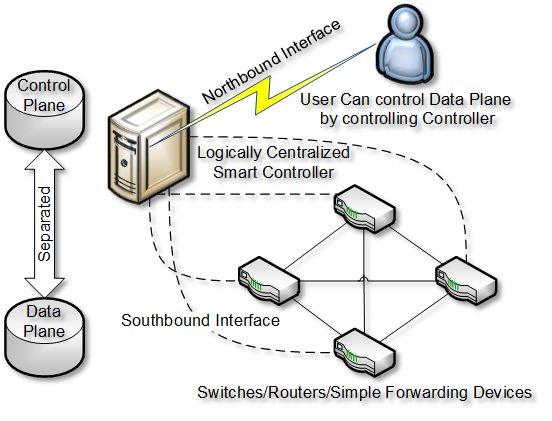}
\caption{\label{simple_sdn_architecture}A simple SDN Architecture}
\end{figure}  

\begin{itemize}

    % \item   \textbf{Forwarding devices} have certain direction sets used to take activities on the approaching. These devices take the task of sending flow or packets to the destination in some logical way. There are some particular types of guidelines which are defined by southbound interfaces. The southbound interface is responsible for connecting between control and data plane. There are different types of southbound interface such as OpenFlow \cite{McKeown.2008}, ForCES \cite{A.Doriaetal.2010}, protocol oblivious forwarding (POF) \cite{Song2014}, which is introduced in the forwarding devices by the SDN controllers executing the southbound protocols.

    \item   \textbf{Data plane} is interconnected devices through wired or non-wired link, and it is designed by network administrator for each mechanism work. In network foundation, interconnected forwarding devices ask for request to the data plane directly to exchange the information. Forwarding devices have certain direction sets used to take activities. These devices take the task of sending flow or packets to the destination in some logical way. There are some particular types of guidelines which are defined by southbound interfaces. The southbound interface is responsible for connecting between control and data plane. There are different types of southbound interface such as OpenFlow \cite{McKeown.2008}, ForCES \cite{A.Doriaetal.2010}, protocol oblivious forwarding (POF) \cite{Song2014}, which is introduced in the forwarding devices by the SDN controllers executing the southbound protocols.

    \item   \textbf{Southbound Interface} is interconnected between control plane and data plane. The network foundation includes the interconnected forwarding devices communicate with the data plane, which is the place where network administrator designs every rules and mechanisms.

    \item   \textbf{Control plane} controls forwarding devices. It can be viewed as the brain of the network, which works to control each component of the whole network architecture. The control logic exists in the applications and controllers which builds up the complete control plane.

    \item   \textbf{Northbound Interface} is the communication between the management plane and control plane. The NOS can offer an Application Programming Interface (API) communicate with an exemplary interface for creating applications to network applications engineers. Usually, a northbound interface abstracts the low-level direction sets utilized by southbound interfaces to program forwarding devices. 

    \item   \textbf{Management plane} is the administrative plane to utilize the use of the capacities offered by the northbound interface to actualize network control and operation logic, where all routing, switching are being done here. In addition, it can be automatically translated to the data plane through the control plane. Besides, northbound and southbound interfaces are the communication link of SDN environment.

\end{itemize}

\subsection{Overview of OpenFlow}

There are the design of southbound API and some other components of SDN \cite{Nunes.2014,Lara.2014,S.T.Ali.2015}, where OpenFlow is the main significant progress. This protocol proposes the communication to interact the data plane with the control of the controller. In \cite{Jarraya.2014}, Jarraya et al. develops a scientific classification for SDN, which allows to control switches, routers remotely by using a network operating system. Usually, NOS is the one who controls the network, and OpenFlow gives a simple stage to follow some rules for forwarding devices. 

On the one hand, OpenFlow allows to the match-action page in each switch/router, where it will make command to take the action if a new packet is matched to the router table. Otherwise, it sends the packet to the controller to check whether the packets need to be forwarded or dropped. This action of the OpenFlow depends on the decision made by the controller, which may set a default rule for the new packets. There are certain amounts of actions, which can be taken such as forwarding, encapsulation, dropping, checking, and matching, and the centralized controller takes all these actions.

\begin{figure}[t!]
\centering
\includegraphics[width=10cm]{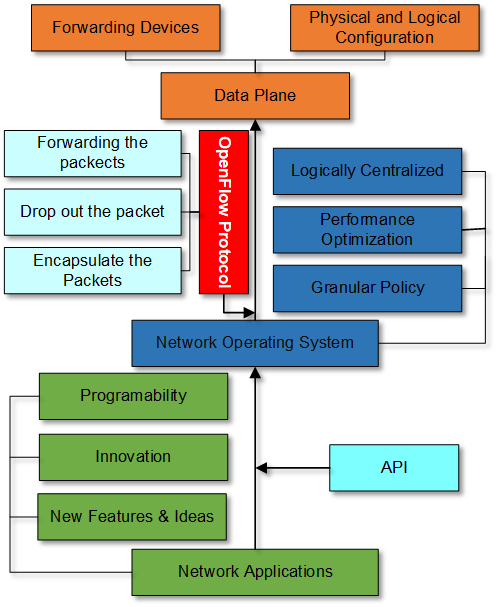}
\caption{\label{openflow}Impact of OpenFlow Protocol in SDN}
\end{figure}

As shown in Fig.~\ref{openflow}, it is clearly visible that how OpenFlow makes the impact during the data processing in SDN environment. Networking applications are implemented in the data plane by networking operating systems and OpenFlow protocol.
% Also, OpenFlow is compatible with the existing network appeals to the interests of researchers and organizations in coming up with some new ideas related SDN. 
SDN traffic measurement depends on such kind of protocol to evaluate the traffic signals. A few OpenFlow-empowered forwarding devices are accessible as business and open source items. With a vast number of flow entries, the vast majority of the switches accessible generally have little ternary content addressable memory (TCAM). It is a reasonable sign that the measurement of the flow tables is developing at a pace intending to address the issues of future SDN traffic measurements. In short, OpenFlow protocol makes the SDN network robust and reliable.
\section{CELLULAR NETWORK}

Even though there has been a consistently developing for constant accessible network \cite{GSMAssociation.2015,Ogul2013}, the security, threats and scalability in cellular networks still need to be researched. Currently existed networks have many security issues such as interface flooding, network element crashing, traffic eavesdropping, unauthorized data access, traffic modification, data modification, compromised network element, malicious insider, theft of service \cite{S.Mavoungou.2016}. These attacks are generally occurred in the base station, radio station and evolved packet core, and it is possible to maximize the security and adequately manage the networks by including SDN in the network. In this part, we discuss the structure of the currently existing network and SDCN architecture respectively.

The tremendous development in the innovation and network is forming a connected universe of millions and billions of devices associated and communicating with each other. The present remote advances in 3G/4G are sprouting IP availability and points in giving quicker web association, sight and sound application and a large number of administrations with expanded performance. In light of these lists of quick improvement in the network devices, the network does not adequately take care of the demand of assorted variety and low latency, which is foreseen in the 5G remote network. The 5G fundamentally gives a client-driven availability where various applications are gotten to a speedier pace, at the higher limit and at 1ms latency. The 5G is viewed as an essential instrument for acknowledging Internet of Things (IoT) worldview interfacing billions of devices as it is equipped for supporting machine-to-machine (M2M) correspondence and requiring little to no effort and low battery utilization.

\subsection{Long Term Evolution}

In this subsection, we will first discuss about the LTE architecture, which is the mainly technology in communication network especially in the SDN, consists of E-UTRAN and EPC \cite{haque_wireless_sdn_survey} two parts. The client hardware (UE) interfaces coordinates traffic through a serving gateway (S-GW) over a GPRS Tunneling Protocol (GTP) with a base station. Here, the S-GW must deal with mobility in a client's area and store many states since clients hold their IP tends to when they move. Also, the S-GW sends traffic to the packet data network gateway (P-GW) to uphold nature of-benefit approaches and screens traffic to perform charging for the clients.
This part needs to be discussed with \cite{amin2016auto, sung2011towards}
The P-GW likewise interfaces with the Internet and other cellular data networks and goes about as a firewall that squares undesirable traffic.

With the joint effort of Mobility Management Entity (MME), which has the handover and mobility functions, the P-GW performs to deal with session setup, reconfiguration and additionally portability. For instance, the P-GW sends QoS and other session data to the S-GW in light of a UE's ask for devoted session setup. Like this, the S-GW advances the data to the MME which at that point requests the base station distribute radio resources and build up the association with the UE. As for the hand-off of UE, the source base station sends the hand-off demand to the real base station and exchanges the UE state to the external base station to an affirmation. The real base station likewise tells the MME that the UE has changed cells, and the former base station will discharge resources, which will create new GTP for a new radio base station. 

The Policy Control and Charging Function (PCRF) oversees flow-based charging in the P-GW. Besides, the PCRF likewise gives the QoS approval that chooses how to treat each traffic flow, in light of the client's membership profile. Also, the Home Subscriber Server (HSS) contains membership data for every client and the related MME. To give a more details of illustration, Fig.~\ref{lte_architecture} shows the LTE architecture as evolved packet core (EPC) and Evolved Terrestrial Radio Access Network (E-UTRAN).

\iffalse
\begin{figure}[t!]

  \centering
  \includegraphics[width=0.45\textwidth]{figure/lte_architecture.png}
  \caption{\label{lte_architecture}LTE Architecture}
  
\end{figure} 
\fi
\begin{figure}[t!]
\centering
\includegraphics[width=15cm]{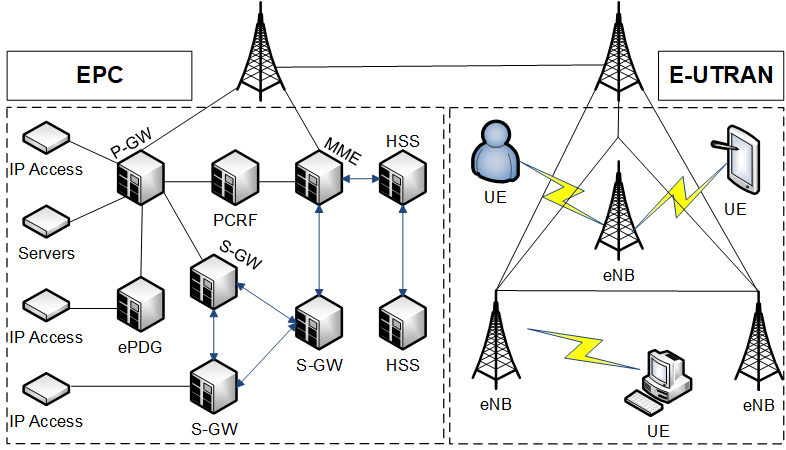}
\caption{\label{lte_architecture}LTE Architecture}
\end{figure} 

In general, the design of present cellular network has a few noteworthy restrictions. All traffic data in the network must pass through P-GW, which makes it a more critical device in EPC AND makes exceptionally costly. Centralized controller data-plane capacities at the mobile Internet limits traffic through the P-GW, including traffic between clients on the same cell network, making it hard to host famous substance inside the cell network. Furthermore, the network devices have vendor-particular design interfaces with no programmability options, impart through complex control-plane protocols, but with a substantial and developing number of tunable parameters. Above all, the new types of architectures are needed to solve these critical problems \cite{haque_wireless_sdn_survey}.

\subsection{Software Defined Cellular Network (SDCN)}

Next in this part, we discuss the structure of SDCN in which Cellular networks are ready for the presentation of Software wireless network \cite{haque_wireless_sdn_survey}, where the network hardware performs fundamental packet-preparing capacities at the command of utilization running on a logically-incorporated controller. As the cellular network is a wireless network, it is not easy to deploy itself to the SDN, which can disperse data-plane rules over various, less expensive network switches, diminishing the scalability on the packet gateway and empowering adaptable treatment of traffic that stays inside the cellular network. Also, it supports ongoing updates to some fine-grained packet handling rules raising enormous scalability challenges. Besides, mobility is a big issue and to solve this problem SDN integration can make a significant impact. Otherwise, it can require forwarding state at the level of individual endorsers, also the state must change rapidly to stay away from benefit disturbances. However, the network state needs to be changed rapidly to perform the network operation.

There are four main ideas mentioned in \cite{haque_wireless_sdn_survey}, and these ideas are flexible policies using local switch agents, flexible switch patterns and remotely control over the virtual-based station. They were proposed to design the cellular network on top of the LTE structure, which also can be used as a 5G network concept. Here, a simple SDCN architecture is shown in Fig.~\ref{sdcn_architecture}. On the other hand, Network Functions Virtualization (NFV)\cite{gharsallahsdn, nguyen2017sdn} is used for specific network functions such as mobility, security, handover, which are stored in the data center, and it also can store user information, billing, subscription. Thus, every logic component is separated from the data plane as different parts. As described in earlier of this paper, the OpenFlow protocol is responsible for the connection between controllers to switches/routers. However, OpenFlow is used to apply the rules directly to the base station from the controller in SDCN, and the controller can be used for directing traffic through middle-boxes, monitoring for network control, billing, mobility, QoS control, virtual cellular operator or inter-cell interference management.
\iffalse
\begin{figure}[t!]

  \centering
  \includegraphics[width=0.45\textwidth]{figure/sdcn_architecture.png}
  \caption{\label{sdcn_architecture}SDCN Architecture}
  
\end{figure}
\fi

\begin{figure}[t!]
\centering
\includegraphics[width=15cm]{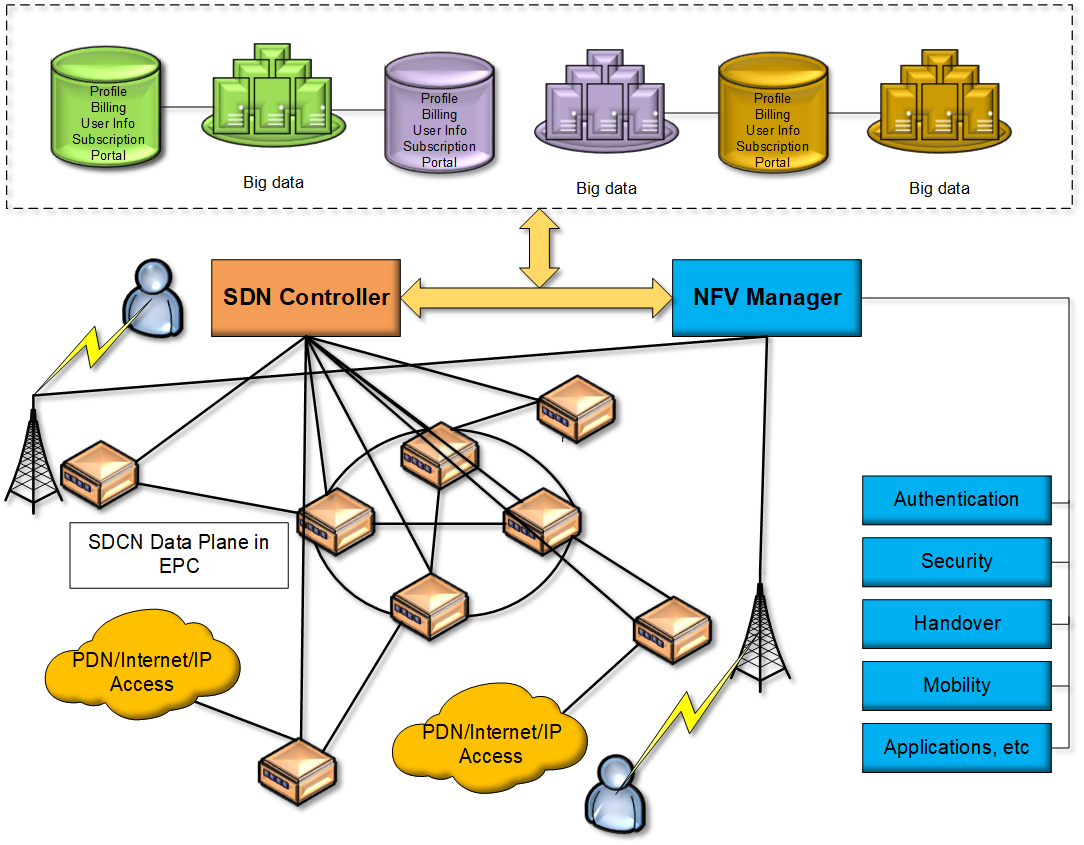}
\caption{\label{sdcn_architecture}SDCN Architecture}
\end{figure} 

\subsection{Security Challenges and Traffic Measurement}

There are four groups of security issues are in the cellular network \cite{S.Mavoungou.2016} of SDCN. Thus, the security challenges are quite relatively with the traffic measurement.
% SDN defeats the confinement of the heterogeneous network by decoupling control plane from the forwarding plane by including programmable capacities, which underscores the way that its division thought furnishes administrators easily of resource provisioning and programmability to change and control the qualities of the entire network \cite{Li.2016}. Thus, network administrators can design dynamic, restrictive software for switches/routers, which can oversee, design, and advance network resources in a snappy and straightforward way. Inspired by this, SDN needs to take care of current network issues and to enhance network security with exceptionally compelling and productive arrangement.
SDN defeats the confinement of the heterogeneous network by decoupling control plane from the forwarding plane by including programmable capacities, which underscores the way that its division thought furnishes administrators easily of resource provisioning and programmability to change and control the qualities of the entire network \cite{Li.2016}.Inspired by this, SDN needs to take care of current network issues and to enhance network security with exceptionally compelling and productive arrangement.

% Fundamentally, switches/routers have poorer performance in equipment resources, which has additionally been viewed as powerlessness in network’s brain \cite{S.T.Ali.2015,Scott-Hayward2013}. Aggressors firstly need to be able to attack the correspondence channel to chop down the connection amongst controller and switches, along these lines, the switches/routers will change into coming up short secure mode or independent mode as indicated by OpenFlow convention. This will significantly influence the network performance. Furthermore, assailants can take over a controller when changes attempted to re-establish association, and will also cause calamities as aggressors have assumed the responsibility of the underlying network devices and control any flows navigated through switches. Although OpenFlow improvement is at the peak, there are still many challenges to be taken into consideration. Especially, the middleman attack is the big issue for SDCN that it is possible to take user information when the data passes through the controller. Thus, an attacker can take network data, change flow rules in the switches, or even embed wrong flow rules to take control of the entire network \cite{Tang.2018}.
Although OpenFlow improvement is at the peak, there are still many challenges to be taken into consideration. Especially, the middleman attack is the big issue for SDCN that it is possible to take user information when the data passes through the controller. Thus, an attacker can take network data, change flow rules in the switches, or even embed wrong flow rules to take control of the entire network \cite{Tang.2018}

Moreover, OpenFlow itself creates some security issues and it will take SDCN into the next level of challenges to solve security issues. The controller generally has a few modules for proficient network administration and observing, that can be viewed as outsider applications \cite{ChuYuHunag.2010}. Once these modules have been endangered in the controller, it will cause unfavorable issue and actuate weakness unexpectedly to the entire framework.

By this way, to overcome these security challenges in SDCN and to set up a superior and shrewder network guard system, we have to apply current developed SDN to recognize them in different tasks and then react it to malevolent activities ahead of time. Here, we believe SDN traffic measurement ought to be the prime decision for the accompanying reasons:

\begin{itemize}
    
    \item   Traffic measurement adopts favorable circumstances of some specific strategies to comprehend and evaluate network practices, which can be extremely useful in identifying mysterious practices ahead of time.
    
    \item   Traffic measurement can likewise be exceptionally helpful in making safeguards and responses in a short time.
    
    \item   Traffic measurement is the way to understand the network status continuously, which can be connected to large network, where these are altogether viable approaches to build up a superior secured environment.
    
\end{itemize}

\section{SDN Traffic Measurement}

In this section, we introduce the traffic measurement in SDN and describe two types of measurements. There are different issues made by varies of the produced movements from current networks. New technology makes troublesome for network administrators to measure the status and progression of the network in compelling and proficient ways \cite{Tso.2013b}, 
% which is a fundamental requirement for the capacity to quantify distinctive kinds of network activity at various time-scales for assignments. 
which is the basic requirement of capacity of clarifying kinds of networks activities at various time-scales.
Congestion control and guaranteed performance are essential to ensure application execution that makes network administrators in extraordinary distressing circumstances to fulfill clients' desires for conveying applications with ensured QoS. There are some key points need to be considered before choosing the way of traffic measurement as following.

\begin{itemize}
    
    \item   Congestion control is the crucial part in the video streaming \cite{Javadtalab.2015}, live video capture and live video chat. We have to precisely identify and measure changes in bandwidth to decrease quality corruption of the video.
    
    \item   Switches/routers are rigid, and it can not manage diverse kinds of network activity, because of the hidden hardwired execution of routing rules \cite{Kim.2013} and various snags. Accurate measurement and correctly programmed in networking will have a chance to mitigate this types of snags, where SDNs and new correspondence protocols like OpenFlow are utilized to solve those issues.
    
    \item   OpenFlow protocol is the key to design a centralized controller for SDN network. By this way, the Network administrator can measure traffic information and make response so that it can give a higher performance in high traffic load
    
\end{itemize}

So far, with the describe of its colossal scale and the assorted variety of the considerable traffic, the SDN measurement tasks need a different way to measure the data traffic. In this way, there are two types of measurement existing in SDN traffic measurement. One is the active measurement, and another is passive measurement \cite{V.MohanY.R.J.ReddyandK.Kaplan.2011}.

\subsection{Active Measurement}

In active measurement process, network flows are ceaselessly checked for execution by sending unique test packets over the network. Also, traffic flows are irritated by test traffic flows, and it may create huge overhead. Thus, active measurement in SDN requires cautious intending to adapt to the necessities of unified control design. Besides, the deployment of active measurement devices expands the data obtaining dramatically by different requests of size \cite{Sezer.2013}. So the flows of data procured by the dispersed devices in the network cannot furnish the SDN controller with the essential data, and it is essential to limit the effect of activity unsettling influence. Even though there are some design methods included in this area like artificial intelligence, robotic nature, the controller faces correspondence bottlenecks, firm control, enhancement issues, and performance optimization, it is necessary to know that this is a continuous process, that the more we run the query and the more congestion will increase. This technique can be vulnerable to the controller and it cannot respond as quickly as the increasing traffic.

\subsection{Passive Measurement}

On the contrary, in passive measurement, genuine traffic is captured and then running queries to it to get the traffic measurement. Passive measurement has no impact as it does not lead to any overhead in the network. However, sometimes passive measurement is vital in substantial traffic instances, as it does not create additional traffic in the network and depends on sampling and some statistical method to get the traffic data. Thus, the primary challenge of passive measurement is that small flows might be missed along an SDN flow way may test the very same packet prompting measurement mistakes \cite{M.JarchelT.ZinnerT.HohnandP.TranGia.2013}. 
% Besides, it also requires advanced diagnostic devices to process network activity as fast as on account of the present data center networks. 
To get a real-time measurement, we need sophisticated hardware to implement with the network so that the result comes out within a tight time. Presently, a few rising methodologies are endeavoring to overcome some critical challenges by effectively and by utilizing the concept of scalability in SDNs to offer programmable interfaces to accomplish fine-grained estimation of network activity flows. These existing investigations are either proposing active measurement techniques or passive measurement strategies. 

\subsection{Requirement of SDN Traffic Measurement}

SDN traffic measurement has some requirements that must be followed to design SDN network. Without accurate measurement the network can not be scalable, adaptable and secured and quite impossible to design higher level network policy \cite{ali2020traffic_Ta_Traffic}. The basic part of SDN is the software program, which is access to be tracked by the attacker. To avoid monitoring what is happening inside the network every time, it is necessary to apply artificial intelligence into the network with the meaningful use of big data. Thus, SDN traffic measurement is the critical step to make it scalable and adaptable for network operators. A robust network framework for SDN has several design goals which will impact the network, and it needs to follow the issues given below.

\begin{itemize}

    \item   \textbf{\textit{Accuracy}}: SDN traffic measurement framework requires high measurement accuracy, which is the crucial part for every network functions specifically in SDN. Besides, a measurement job would not create congestion and performance degradation.

    \item   \textbf{\textit{Resource effectiveness}}: SDN traffic measurement framework proficiently uses Central Processing Unit (CPU) for packet handling and memory on which the measurement job would like to increase extra usage of the resources. That is why CPU usage will be increased by the increasing number of measurement tasks. Thus, the solutions of traffic measurement need to decrease resource usage while maintaining higher accuracy. 

    \item   \textbf{\textit{Generality}}: SDN traffic measurement framework supports an extensive variety of measurement solutions. We can not use a framework only for one network function, which is costly, complex, unreliable and unsuitable for next generation heterogeneous network.

    \item   \textbf{\textit{Simplicity}}: SDN traffic measurement framework naturally mitigates the handling weights under high traffic load, and it does not necessarily require configuring every host. It needs to arrange the data and the results to deploy an action immediately in the network.

\end{itemize}

Some of the papers related to SDN traffic measurement framework are the review of outline such as OpenSketch \cite{MinlanYu.}, Univmon \cite{Liu.}, Dream \cite{Moshref.2014}, Scream \cite{Moshref.2015}, Trumpet \cite{Moshref.2016}, and online measurement of large traffic data by Jose et al. \cite{jose_online_measurement_traffic}, where all of them supports these requirements above. It contains a separated data plane that keeps running on the software switches in a network and a centralized control plane that collects data from all software switches for various types of measurements. Fig.~\ref{req_sdn_meas} shows balance of overhead, generality, real-time decision making, and proper resource usage are the standards set for SDN accurate measurement.

\iffalse
\begin{figure}[t!]
  \centering
  \includegraphics[width=0.45\textwidth]{figure/req_sdn_meas.png}
  \caption{\label{req_sdn_meas}\Centering Requirement of SDN traffic measurement}
\end{figure}
\fi

\begin{figure}[t!]
\centering
\includegraphics[width=10cm]{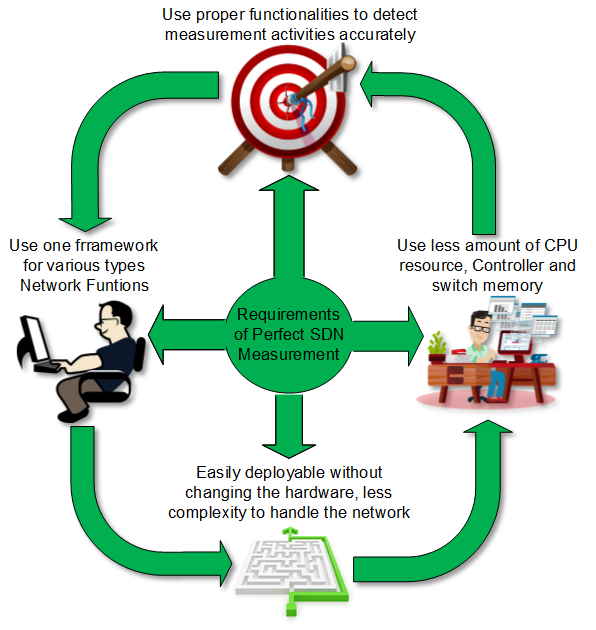}
\caption{\label{req_sdn_meas}Requirement of SDN traffic measurement}
\end{figure}

\subsection{Traffic Measurement Activities}

There are several measurement activities for traffic measurement. Here, we have to find out which in the network needs to be measured. To answer this question, we summarize the measurement activities, which is inevitable for SDN/SDCN.

\subsubsection{\textbf{Link Latency}} Link latency is the delay from input into a system to desired outcome; the term is understood slightly differently in various contexts and latency issues also vary from one system to another. It greatly affects how usable and enjoyable electronic and mechanical devices as well as communications are, and it is sensitive to network applications to find out the malfunctions of the network and then correct it \cite{McKeown.2008}. For example, e-banking needs to communicate without compromising the delay where link latency plays a crucial rule. By watching latency thresholds, the network administrator can re-route the network path to maintain QoS \cite{A.Doriaetal.2010}.

There is an example of the link latency and how we can measure it as shown in Fig.~\ref{latency}. In general, it can be described as the latency in the communication between the user and the servers. Here, we need to calculate the latency between them as one part from pulling request to receiving instruction, then we can make solution to guarantee the QoS.

\iffalse
\begin{figure*}[ht!]
\begin{adjustwidth}{-\extralength}{0cm}
    \centering
    {\includegraphics[width=0.55\textwidth]{figure/link_latency.png}\label{latency}}
    \subfloat[An Example of Link-Latency]
    \hspace{15pt}
   {\includegraphics[width=0.35\textwidth]{figure/network_topology.png}\label{topology}}
    \subfloat[An Example of Network Topology]
    \caption{The example of Link-Latency and Network Topology in SDN Measurement Activities. (a) refers to the connect latency to send requests between host and server, and (b) shows a distributed topology in a simple network.}
\end{adjustwidth}
\end{figure*}
\fi

\begin{figure}[!b]
\centering
\begin{subfigure}[b]{0.5\columnwidth}
    \centering
    \includegraphics[width=\columnwidth]{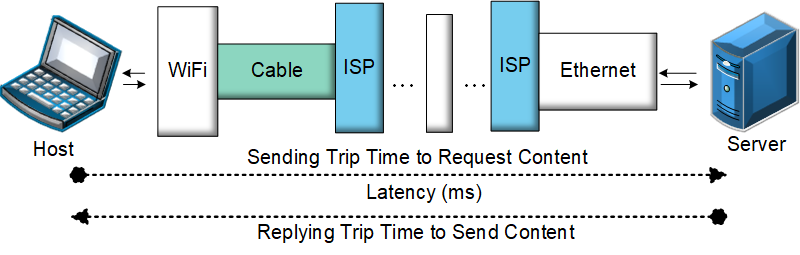}
    \caption{\centering An Example of Link-Latency}
    \label{latency}
\end{subfigure}
\hspace{15pt}
    \begin{subfigure}[b]{0.45\columnwidth}
    \centering
    \includegraphics[width=\columnwidth]{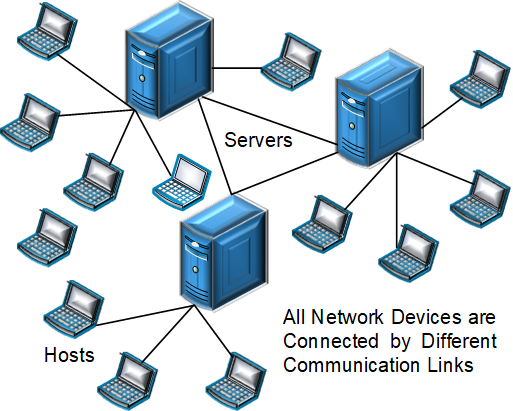}
    \caption{\centering An Example of Network Topology}
    \label{topology}
\end{subfigure}
\caption{The example of Link-Latency and Network Topology in SDN Measurement Activities. (a) refers to the connect latency to send requests between host and server, and (b) shows a distributed topology in a simple network.}
\end{figure}

\subsubsection{\textbf{Network Topology}} Network topology is the arrangement of the elements like links or nodes of a communication network, which is the complete overview of the whole network contains all the physical connections link among all network nodes \cite{Bakshi2013,Alhanani2014,nehra2018tilak}.
Estimating and refreshing topology assume an essential job in giving capacities of essential network. What is more, some network functions such as routing, troubleshooting, network management, malware detection need the information of network topology.

Fig.~\ref{topology} shows a distributed topology in a simple network, the nodes contact to each other and each of them links to several users. To measure the traffic in different kinds of topology networks, we need to first clarify their architecture and then collect the messages.
\iffalse
\begin{figure*}[ht!]
\centering
\begin{minipage}[t]{0.47\textwidth}
\centering
\includegraphics{figure/bandwidth.png}\label{bandwidth}
\caption{An Example of Bandwidth}
\end{minipage}
\hfill
\begin{minipage}[t]{0.47\textwidth}
\centering
\includegraphics{figure/heavy_hitter.png}\label{heavy_hitter}
\caption{\centering An Example of Heavy Hitter and Heavy Changer}
\end{minipage}
\end{figure*}
\fi

\begin{figure}[t!]
\centering
\begin{subfigure}[b]{0.47\columnwidth}
    \centering
    \includegraphics[width=\columnwidth]{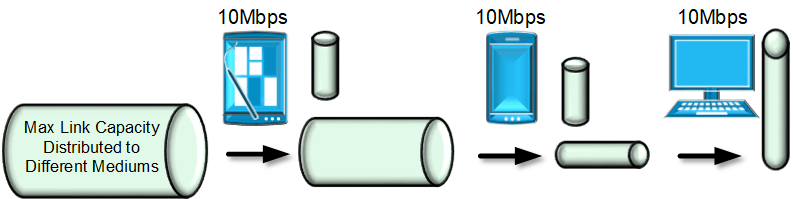}
    \caption{\centering An Example of Bandwidth}\label{bandwidth}
\end{subfigure}
\hspace{15pt}
    \begin{subfigure}[b]{0.47\columnwidth}
    \centering
    \includegraphics[width=\columnwidth]{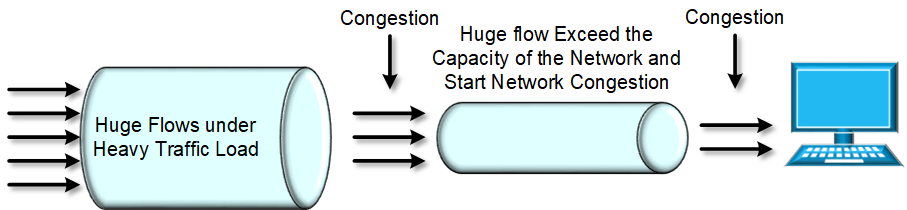}
    \caption{\centering An Example of Heavy Hitter and Heavy Changer}\label{heavy_hitter}
\end{subfigure}
\caption{An Example of Heavy Hitter and Heavy Changer}
\end{figure}

\subsubsection{\textbf{Bandwidth}} Bandwidth \cite{wang_bandwidth_allocation_strategy,paul2016enhanced,megyesi2017challenges} is the maximum throughput or capacity of the communication whether it is logical or physical communication.
Maximum capacity sustains on the shanon channel capacity theorem, which depends on signal to noise ratio. An example is shown in Fig.~\ref{bandwidth}, that there is the limit capacity on the global branch at each second whether it will dredge into several local branches.

\subsubsection{\textbf{Heavy Hitters} and \textbf{Heavy Changer}} Heavy hitters \cite{Zhang2010} are huge flows, which cause network congestion, make link hard handle the data. Thus, we have to set a threshold point to decide whether the traffic is jammed, and then give out the solutions. 
% Due to increasing speeds of the network nowadays, this one is attracted to the researchers.
On the other hand, heavy changer is almost the same as a heavy hitter that also cause the congestion of the network. The difference between them is that the heavy changer is a flow whose difference in byte counts crosswise over two consecutive epochs exceeds a threshold. Fig.~\ref{heavy_hitter} shows the example of these two measurement, where the huge flows exceed the capacity of the network and lead to the congestion.

\iffalse
\begin{figure*}[ht!]
    \centering
    \subfloat[An Example of DDoS Attack in SDN]{\includegraphics[width=0.5\textwidth]{figure/ddos.png}\label{ddos}}
    \hspace{15pt}
    \subfloat[An Example of Superspreader in SDN]{\includegraphics[width=0.41\textwidth]{figure/superspreader.png}\label{superspreader}}
    \caption{The example of DDoS and Superspreader Attack in SDN Measurement Activities. The main difference between these two attacks is that whether the victim is used to receive amount of requests or used to send plenty of packet flows.}
\end{figure*}
\fi
\begin{figure}[t!]
\centering
\begin{subfigure}[b]{0.53\columnwidth}
    \centering
    \includegraphics[width=\columnwidth]{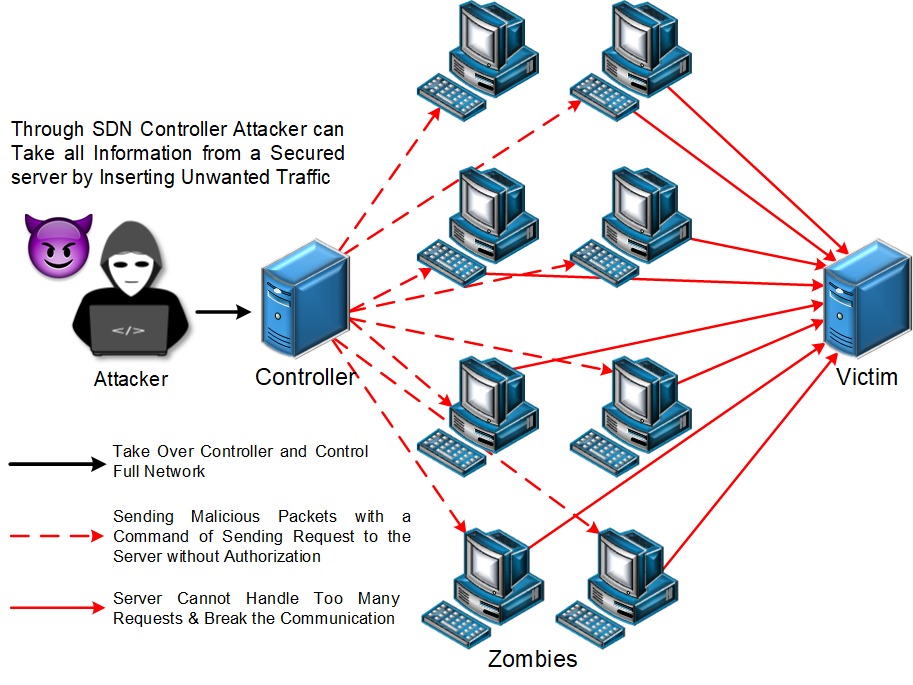}
    \caption{An Example of DDoS Attack in SDN}
    \label{ddos}
\end{subfigure}
\hspace{15pt}
    \begin{subfigure}[b]{0.42\columnwidth}
    \centering
    \includegraphics[width=\columnwidth]{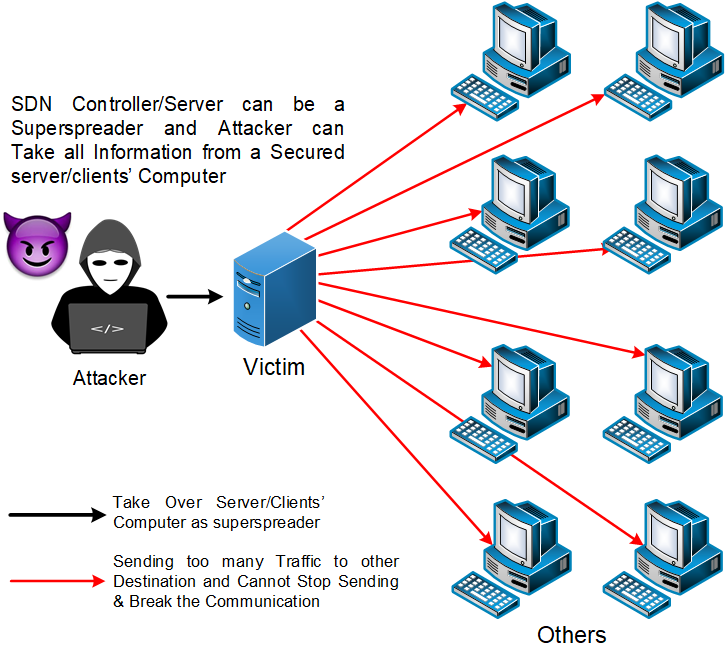}
    \caption{An Example of Superspreader in SDN}
    \label{superspreader}
\end{subfigure}
\caption{The example of DDoS and Superspreader Attack in SDN Measurement Activities. The main difference between these two attacks is that whether the victim is used to receive amount of requests or used to send plenty of packet flows.}
\end{figure}

\subsubsection{\textbf{DDoS}} DDoS \cite{Nagpal2015} stands for Distributed Denial of Service where a system gets a massive amount of traffic measurement more than a threshold point from various sources, that leads to break down the network; and we called this DDoS attack. Thus, the server is attacked by amount of disturbing flows, and then the link cannot handle the traffic and stop responding to all requests. DDos attack is one of the popular attacks which takes less time to attack and control over the network.

As shown in Fig.~\ref{ddos}, the attacker uses the controller to control plenty of zombie services to send traffic massages to the victim, which produces the congestion and breaks down the victim system. In the real-time, our measurement needs to filter out the attack packets to provide QoS.

\subsubsection{\textbf{Superspreader}} Superspreader \cite{Kamiyama2007} is opposite to DDoS, that amount of flows/data are sent by a source more than threshold point to several destinations. Thus, an attacker can use one client’s computer as a superspreader to send much traffic to other destinations, which leads the client’s computer cannot handle the traffic and even not stop responding. To avoid being attacked by the superspreader, we must disconnect the internet of whole network. Different from the DDoS, the attacker in Fig.~\ref{superspreader} controls the victim computer to send amount of flows to other places at each time, and victim cannot handle this traffic as well. 

\subsubsection{\textbf{Entropy}} Entropy \cite{SHI2014} is the equilibrium state of the network. There are many links in the network, which can give the largest number of ways to construct the network. Thus, a network needs to use all links properly with the real time scenario. It is crucial to know the link flow distribution so that the link can be at the equilibrium state.

There are some other crucial measurement activities such as, cardinality (number of particular flows in an epoch), flow estimate distribution (flows for ranges scopes of byte counts in an epoch), also are valuable to be measured in the traffic measurement task.
%%%%%%%%%%%%%%%%%%%%%%%%%%%%%%%%%%%%%

\section{SDN Traffic Measurement Solutions}

In this section, we will discuss existing proposed SDN traffic measurement solutions, which can be used for further research to design secured SDCN architecture. We collect almost all the papers which proposed traffic measurement solutions until 2018, from where the initial stage network architecture was simple, less complex and easy to deploy. Thus, in the research before, measurement requirement was based on overhead and most of the researchers concentrated on balancing overhead. At the same time, resource usage policy played an crucial rule and some other solutions were introduced to get high accuracy by using less resources. For the complex and heterogeneous of today's network, we need to instant video calling \cite{salman2020link_UOB}, online transaction, social networking, and we have to consider about real time traffic measurements for quick decision making. Furthermore, we need some specific network discovery such as latency monitoring, topology discovery to find out the current status of the network. Here, we make 4 groups of different traffic measurement solutions in Fig.~\ref{sdn_references}, which will be given below in details.
\iffalse
\begin{figure*}
  
    \centering
    \includegraphics[width=17cm]{figure/meas_ref.png}
    \caption{\label{sdn_references}References of SDN Traffic measurement}
\end{figure*}
\fi

\begin{figure}[t!]
\centering
\includegraphics[width=16cm]{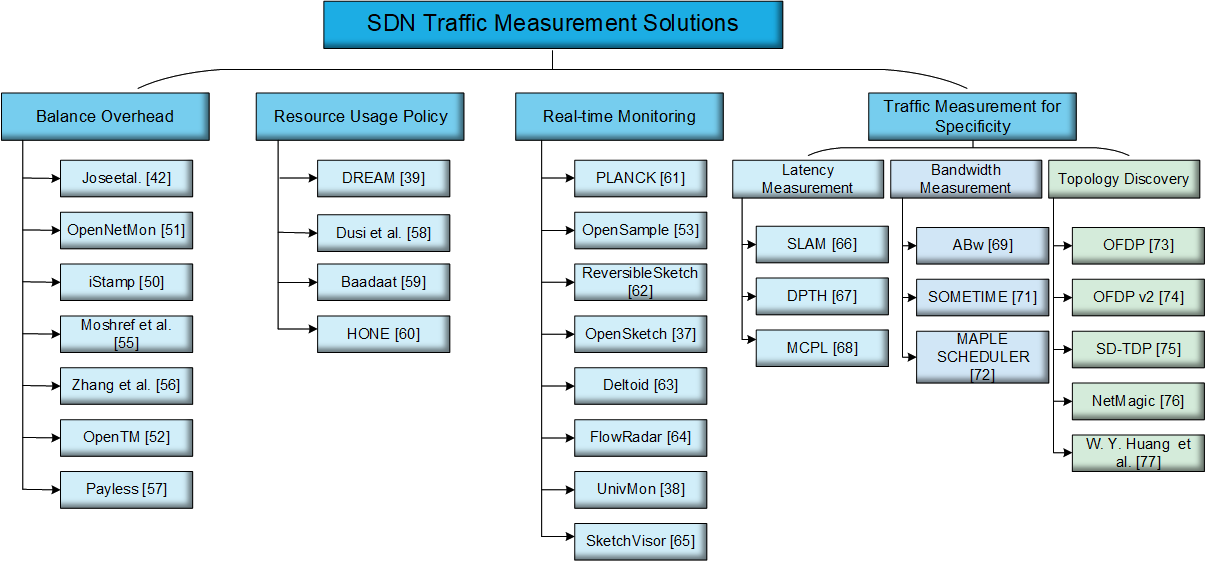}
\caption{\label{sdn_references}References of SDN Traffic measurement}
\end{figure}  

\subsection{Balance Overhead}

Ceaselessly observing the network frequently presents overhead, should be taken into consideration as a trade-off with traffic measurement precision. With an end goal to locate a suitable zone amongst exactness and overhead ramifications, Jose \textit{et al.} \cite{jose_online_measurement_traffic} proposed a way to collect substantial traffic data in commodity switches by measurement structure where switches coordinate packets against a collection of wildcard case rules accessible in Ternary Content Addressable Memory (TCAM). Here, TCAM is a type of RAM where data can be saved in boolean form, and it can save a good amount of data by compression. This approach reduces overhead because the switch can take decisions based on some wildcard rules. By using TCAM, it is possible to store data or packet processing rules in the switch/router. The structure is assessed utilizing a Hierarchical Heavy Hitter (HHH) program to comprehend the trade-off amongst accuracy and overhead. In these types of the category, we need to update regularly the matching rules, which are the primary concern of this measurement solution \cite{T_SDN_DC}.

To further improve the above method, \textit{iSTAMP} \cite{Malboubi.2014} powerfully segments the TCAM entries to permit fine-grained measurement tasks of coming flows. iSTAMP creates two partitions of TCAM, one is used for aggregation, and the other is for deaggregation. Flows are stamped for active measurement on the off chance that they are considered to be essential. The iSTAMP utilizes the algorithm which is based on calculation to process these two arrangements of measurements, that are then mutually prepared to evaluate the extent of all network flows utilizing distinctive optimization strategies.

On the other hand, methods like \textit{OpenNetMon} \cite{vanAdrichem.2014} used OpenFlow to quantify traffic parameters. OpenNetMon decides whether end-to-end QoS parameters are exists in each flow,  and it is a process of continuously monitored with predefined rules. Thus this active measurement fetches the data from switches, and the queries will vary by changing the flow rate. Also if flow rate changes, the query will increase and vice-versa. Another OpenFlow based approach is proposed by \textit{OpenTM} \cite{Tootoonchian.2010}. Switches are simple forwarding devices here, and the controller can query traffic data by using OpenFlow flow entries, of which the logic is to track every active flow in the network. OpenTM is an active far-reaching measurement approach that at last will present overhead during the time spent intermittently pulling factual data from switches over the network. Besides, OpenTM is used to mix of determination strategies to choose switches for pulling data; this may prompt some measurement mistake. Thus, OpenTM is not suitable for high traffic congestion.

There are also many approaches in different method. \textit{Hash-based switches} \cite{Moshref.2013} are used to collect the traffic flows in the SDN network, which helps distinctive measurement undertakings with the heavy hitter calculation for characterizing vital traffic. In any case, observing guidelines should be carefully designated over the network. Like in Y. Zhang\textit{ et al.} \cite{Zhang.2013}, it utilizes a forecast based calculation for flow checking to distinguish anomalies. The measurement granularity along both the spatial and the worldly measurements changes progressively. As the same, abnormality locators can educate the flow-gathering module to give fine-grained measurement data if there should arise an occurrence of expecting an attack or it can gather fine-grained flow data generally. Furthermore, \textit{Payless} \cite{Chowdhury.2014} is an SDN based active framework, which is also considered with its accuracy and overhead. The preferred fundamental standpoint of Payless is that it utilizes versatile measurements gathering calculation to achieve exact data continuously without bringing about noteworthy network overhead. The API of floodlight controller is used for actualization, which has low overhead and makes higher precision. Table \ref{table_balanced_overhead}, shows that comparison of the traffic measurement by considering the balanced overhead.

From the approaches mentioned above, we can generalize their characteristics as following:

\begin{itemize}
    
    \item   These traffic measurement solutions are almost concentrated on balancing overhead and the accuracy.
    \item   TCAM is the vital component of all these solutions in this part, and flow entries are stored in the different segment of TCAM.
    \item   A different approach of by collecting TCP sequence number in sampled flow (sFlow) collector makes it robust in hierarchical network and popular among the network operators.

\end{itemize}

\begin{table*}[ht!]
\small
\caption{SDN Traffic Measurement Solutions Considering Balanced Overhead}
\label{table_balanced_overhead}

    \centering
    \begin{tabular}{p{0.15\linewidth}p{0.37\linewidth}p{0.37\linewidth}} \hline
	\textbf{Reference} & \textbf{Solutions} & \textbf{Drawbacks} \\ \hline
	
    \textit{Jose et al. \cite{jose_online_measurement_traffic}} & 
    
    \begin{itemize}
        
        \vspace{-0.15cm}
        \item   Coordinating switch packets against a collection of wildcard rules
        \item   Taking decisions of switches by using wildcard rules using Ternary Content Addressable Memory (TCAM)
        \vspace{-0.15cm}
        
    \end{itemize}   &
    
    \begin{itemize}
        
        \vspace{-0.15cm}
        \item   Continuously observing network flows may lead overhead
        \item   Switch does not have a lot of memory spaces
        \item   Updating rules by matching with wildcard entries is computationally costly 
        \vspace{-0.15cm}
        
    \end{itemize}   \\  \hline
    
    \textit{OpenNetMon \cite{vanAdrichem.2014}} & 
    
    \begin{itemize}
        
        \vspace{-0.15cm}
        \item   Continuously monitoring of the network flows with predefined rules
        \item   Fetching the data from switches using OpenFlow protocol
        \vspace{-0.15cm}
        
    \end{itemize}   &
    
    \begin{itemize}
    
        \vspace{-0.15cm}
        \item   Changing flow rate can increase flow rate
        \item   Increasing flow rate may lead to overhead
        \vspace{-0.15cm}
        
    \end{itemize}   \\  \hline
    
    \textit{iStamp \cite{Malboubi.2014}} & 
    
    \begin{itemize}
        
        \vspace{-0.15cm}
        \item   Creating two partitions of TCAM using OpenFlow control
        \item   Flows are stamped when it is essential
        \item   Segregating two arrangements of algorithm, aggregation, and de-aggregation
        \vspace{-0.15cm}
        
    \end{itemize}   &
    
    \begin{itemize}
        
        \vspace{-0.15cm}
        \item   Introducing additional technique called STAMP, which is computationally costly
        \item   It needs the crucial flows to get higher accuracy
        \vspace{-0.15cm}
        
    \end{itemize}   \\  \hline
    
    \textit{Moshref et al. \cite{Moshref.2013}} &
    
    \begin{itemize}
        
        \vspace{-0.15cm}
        \item   Used for gathering traffic data
        \item   Handling of monitoring rules
        \vspace{-0.15cm}
        
    \end{itemize}   &
    
    \begin{itemize}
    
        \vspace{-0.15cm}
        \item    Monitoring rules need to delegates carefully across the network to get higher accuracy
        \vspace{-0.15cm}
        
    \end{itemize}   \\  \hline

    \textit{Zhang \cite{Zhang.2013}} &
    
    \begin{itemize}
    
        \vspace{-0.15cm}
        \item   Forecasting based algorithm for flow counting to detect anomalies
        \vspace{-0.15cm}
        
    \end{itemize}   &
    
    \begin{itemize}
        
        \vspace{-0.15cm}
        \item   Accurate only for identified traffic
        \vspace{-0.15cm}
        
    \end{itemize}   \\  \hline

    \textit{OpenTM \cite{Tootoonchian.2010}} &
    
    \begin{itemize}
        
        \vspace{-0.15cm}
        \item   Monitoring every active flow and pulls data continuously from switches
        \item   Giving high accuracy by continuously monitoring
        \vspace{-0.15cm}
        
    \end{itemize}   &
    
    \begin{itemize}
    
        \vspace{-0.15cm}
        \item   May lead to high overhead after a certain limit
        \item   Measurement mistakes can be occurred due to determine strategies to choose switches for pulling data
        \vspace{-0.15cm}
        
    \end{itemize}   \\  \hline

    \textit{Payless \cite{Chowdhury.2014}} &
    
        \begin{itemize}
        
        \vspace{-0.15cm}
        \item   Polling flow statistics with a regular interval
        \item   Concentrating tradeoff between accuracy and overhead
        \vspace{-0.15cm}
        
    \end{itemize}   &
    
    \begin{itemize}
        
        \vspace{-0.15cm}
        \item   Accuracy depends on based on the length of the polling interval
        \item    Finding a suitable polling interval is difficult
        \vspace{-0.15cm}
        
    \end{itemize}   \\  \hline

    \end{tabular}
\end{table*}

\subsection{Resource Usage Policy}

As we mentioned before, resource usage was not a concerning trade-off for above solutions; so a proposed method \textit{DREAM} \cite{Moshref.2014} is a dynamic resource allocation software-defined measurement system that harmonies between accuracy levels and resources are used for measurement activities. These resources are not allotted before the execution of the measurement task in DREAM, which concerns resource usage and its impact on measurement accuracy, to get an accurate measurement. Furthermore, DREAM system tries to utilize heavy hitter programs demonstrate that DREAM can bolster more simultaneous undertakings with higher accuracy than a few different options. Also, alignment is a necessary component between source and destination so that it can achieve a higher-level accuracy with reducing overhead.

Also, another way is developed by Dusi \textit{et al.} \cite{Dusi.2014} that proposed a powerful proactive controller, which required a certain amount of space for flow entries in TCAM. However, sometimes traffic flow can exceed the limit of TCAM entries. While furnishing SDN switches with all the more capable TCAMs is a possible alternative, this may come to the detriment of expanding the operation and power utilization costs. The investigation recommends that controllers ought to expend resources proficiently utilizing a reactive logic control approach. As in DREAM, the examination recommends that resources must be dispensed and liberated relying upon the network stack, the viable conduct of the flows, their granularity and packet processing. Thus, it should be a way that can be manageable in the network correctly. Besides, \textit{Baatdaat} \cite{Tso.2013} is another proposal, which utilizes NetFPGA programmable switches with running OpenFlow, and it allows continuous dynamic flow. The proposed algorithm can adjust to immediate traffic blasts and additionally to standard connection stack by utilizing save DC network ability to moderate the performance corruption of intensely used connections.

A platform named \textit{HONE} where various types of measurement solutions can be gathered together is proposed in \cite{Sun.2015}. Here, it presents a uniform stack for a various accumulation of measurements in SDN based frameworks. Since consistently gathering measurable data about network flows is costly, two strategies are proposed to solve this issue. The first strategy makes a table of statistical data collected from source, destination and network devices, which minimizes the network overhead by allowing controller for querying from that statistical data table. The second strategy is known as parallel streaming, where operators can use this data by aggregating among multiple hosts. If we want to apply HONE to an SDN network, scalability will be the significant challenges because each host needs to install a software that synchronized to the statistical data so that meaningful queries can be achieved. Table \ref{table_resource_usage}, shows that comparison of the traffic measurement by considering the resource usage.

\begin{table*}[ht!]
\small
\caption{SDN Traffic Measurement Solutions Considering Resource Usage}
\label{table_resource_usage}

    \centering
    \begin{tabular}{p{0.15\linewidth}p{0.37\linewidth}p{0.37\linewidth}} \hline
	\textbf{Reference} & \textbf{Solutions} & \textbf{Drawbacks} \\ \hline
    
    \textit{DREAM \cite{Moshref.2014}} & 
    
    \begin{itemize}
        
        \vspace{-0.15cm}
        \item   Maintaining harmony between overhead and resource usage as resources are not allotted before measurement tasks
        \item   Deploying resources dynamically to get desired level accuracy
        \vspace{-0.15cm}
        
    \end{itemize}   &
    
    \begin{itemize}
        
        \vspace{-0.15cm}
        \item   Alignment is necessary between source and destination to get higher accuracy
        \vspace{-0.15cm}
        
    \end{itemize}   \\  \hline

    \textit{Dusi et al. \cite{Dusi.2014}} &
    
    \begin{itemize}
        
        \vspace{-0.15cm}
        \item   Intending to proactive using a Certain amount of space for flow entries in TCAM
        \item   Controlling the TCAM space
        \item   Powerful TCAM is used for increasing the capacity of flow entries
        \vspace{-0.15cm}
        
    \end{itemize}   &
    
    \begin{itemize}
        
        \vspace{-0.15cm}
        \item   More power usage and computationally costly
        \item   Accuracy increased by adding powerful TCAM
        \vspace{-0.15cm}
        
    \end{itemize}   \\  \hline

    \textit{Baadaat \cite{Tso.2013}} &
    
    \begin{itemize}
        
        \vspace{-0.15cm}
        \item   Utilizing OpenFlow, and NetFPGA programmable switch
        \item   Allowing the switch continuously for dynamic flow scheduling
        \vspace{-0.15cm}
        
    \end{itemize}   &
    
    \begin{itemize}
        
        \vspace{-0.15cm}
        \item   Immediately traffic bursts do not adapt link load
        \vspace{-0.15cm}
        
    \end{itemize}   \\  \hline

    \textit{HONE \cite{Sun.2015}} &
    
    \begin{itemize}
        
        \vspace{-0.15cm}
        \item   Residing software agent on every host
        \item   Interacting module with network devices
        \item   Collecting statistical data periodically
        \item   Following two strategies; one is to make a table of statistical data collected from source and second one is parallel streaming
        \vspace{-0.15cm}
        
    \end{itemize}   &
    
    \begin{itemize}
        
        \vspace{-0.15cm}
        \item   Continuously collecting data is costly
        \item   Serious scalability problem due to reside software agents in every host
        \vspace{-0.15cm}
        
    \end{itemize}   \\  \hline

    \end{tabular}
\end{table*}

From the approaches mentioned above, we can generalize their characteristics as following:

\begin{itemize}

    \item   These all traffic measurement solutions before concentrated on how to decrease resourse usage of CPU with improving the perfomance of accuracy.
    \item   Powerful proactive controller was used for getting more TCAM storage, but flow entries exceeded the limit of the table in high traffic load. Thus, researchers utilized optimization techniques to manage resources more efficiently.
    \item   One of the accurate technique they applied that, different measurement solutions got different priority on resource usage based on the measurement requirement.

\end{itemize}

\subsection{Real-time Monitoring}

Collecting statistical data of the large amount of flows in real-time is the challenge in SDN traffic measurement. Thus, time-sensitive apps and real-time analysis can solve this issue. A proposed method, \textit{PLANCK} \cite{Rasley.2014} is a real-time measurement framework which gathers significant real-time data for statistical analysis, and it utilizes the typical characteristics of the switches. The eminent advantage is that it can be deployed in an SDN environment without changing network devices. Also, PLANCK gives a speedy network framework for comparing traditional network as port mirroring, which is a typical way to deal with traffic measurement going through the mirror utilizing an assortment of network analyzers and security applications. Nevertheless, traffic volume may surpass the limit of the ports to make the switch begin dropping packets. To unravel this issue, the researchers proposed a way \cite{Suh.2014} to buffer the traffic measurements for advance examination. However, buffering does not give proper solutions because dropping packets are dramatically increased in high network load.

Slightly different from the above OpenFlow approach, \textit{OpenSample} \cite{Suh.2014} is a sampling based measurement method proposed by IBM, a manufacturer company of hardware devices. It uses sFlow \cite{MakingtheNetworkVisible.} packets to give near-continuous measurements of both network load and individual flows by catching packets from the network. OpenSample utilizes TCP sequence number from the captured data, and sFlow collector collects the data to analyze accurate flow statistics. Floodlight OpenFlow controller is used for the testbed. One of the fundamental focal points of OpenSample is that it considers any TCP flow. Furthermore, it can negotiate colossal traffic and continuous flows and no requirement of adjustment to switches. These advantages make exceptionally attractive and acceptable to global network operators. However, this is expensive and vendor oriented to be deployed in SDN.

A new idea of using sketch-based measurement is started from \textit{Reversible Sketch} \cite{Schweller.2007}, which utilizes flow header and hashing function to make it smaller subspace. It takes a less amount of memory space. In the first stage, it records the packet stream using FPGA board by operating it online. Also, it can detect the massive changes of the flows in the next stage. Thus, this framework is critical in heavy traffic, which uses a little amount of memory. Furthermore, most sketch-based arrangements are intended for the particular measurement task. To run various measurement tasks together, we have to use a set of algorithm altogether to get the desired query. In addition, running every algorithm individually on each packet turns out to be computationally demanding; so we need a framework for measurement solutions. 

A complete framework, \textit{OpenSketch} \cite{MinlanYu.} requires to sketch the data with three stages of hashing, classification, and counting. The sketches used in it are just natives that cannot be straightforwardly utilized for network measurement instead we should supplement them with extra segments and operations to ultimately support a measurement task. Besides, the network administrator uses a measurement library to get the desired measurement tasks and it can use a set of the algorithm from the library to get different types of measurement. OpenSketch can be utilized for a few measurement exercises including Heavy Hitter measurement, traffic flows, DDoS. Specifically, keeping in mind that to collect traffic measurements, adding extensions is necessary. Sketch is a program that is reversible, implying that sketches store traffic statistics, as well as productively answer the queries on the measurements. After querying of flow, Count-Min sketch algorithm \cite{MinlanYu.} can return a flow size. Thus, we need to distinguish heavy hitters that surpass a pre-determined threshold; a Count-Min sketch, which can instantly report a heavy hitter by checking a packet whether it surpasses the limit. Nevertheless, the earlier threshold is inaccessible ahead of time, and we have to inquire for heavy hitters subject to various thresholds. In addition, we should query all flows in the whole flow space and check if each of them surpasses the threshold for the possibility of giving higher accuracy in measurement tasks. Above all OpenSketch has many advantages, which almost overcome all other problems. Some of the essential benefit of using this one as follows:

\begin{itemize}
    
    \item   Using hashing and only count the flows increases the accuracy with low overhead.
    \item   Utilizing a very less amount of resources. 
    \item   Having a measurement library for measurement reduces computational cost and time.
    
\end{itemize}

\textit{Deltoid} \cite{Cormode.2005} updates every packet and adds counters in each bucket which is encoded in flow headers. This updating significantly increases the query cost and time-consuming. However, it is not easy to get back the data from the data plane. Another solution \textit{FlowRadar} \cite{Li.2016b} solves this issue by mapping flows to counters through exclusive or (XOR) operations, where flows can be constructed easily by repeatedly XOR-ing. However, query cost and heavy computational overhead is still the basic challenge. Recently, proposed \textit{UnivMon} \cite{Liu.} enables to sketch the data simultaneously with distinctive sorts of traffic measurements. Nonetheless, it needs to refresh different components and remains computationally costly. After UnivMon there is another paper \textit{SketchVisor} \cite{Huang2017} proposed a new idea of the faster path. Generally in colossal traffic scenario measurement accuracy in other solutions are computationally high costly and less accurate. SketchVisor framework introduces a faster pathway to bypass the normal path, when network congestion is increased. It only counts faster path flows and small flows count can be achieved by subtracting fast path flows count from total flows count. Although sketch based solutions are attractive in SDN traffic measurement or arrangement, it needs to be updated network nodes continuously. Table \ref{table_real_time_monitoring}, shows that comparison of the traffic measurement by considering the balanced overhead.
 
From the above methods, we can make a summarize as following:
\begin{table*}[htp!]
\small
\caption{SDN Traffic Measurement Solutions Considering Real-time Monitoring}
\label{table_real_time_monitoring}

    \centering
    \begin{tabular}{p{0.15\linewidth}p{0.37\linewidth}p{0.37\linewidth}} \hline
	\textbf{Reference} & \textbf{Solutions} & \textbf{Drawbacks} \\ \hline

    \textit{PLANCK \cite{Rasley.2014}} &
    
    \begin{itemize}
    
        \vspace{-0.15cm}
        \item   Utilizing port mirroring that exists all commodity switches by deployed without changing network devices
        \item   Collecting statistical data very fastly
        \vspace{-0.15cm}
        
    \end{itemize}   &
    
    \begin{itemize}
        
        \vspace{-0.15cm}
        \item   Traffic can exceeding traffic limit the capacity of the ports
        \item   Packets continuously drop in high buffering
        \vspace{-0.15cm}
        
    \end{itemize}   \\  \hline

    \textit{OpenSample \cite{Suh.2014}} &
    
        \begin{itemize}
        
        \vspace{-0.15cm}
        \item   Sampling-based measurement technique
        \item   Utilizeing TCP sequence number from captured data, and sFlow collector collects the data
        \item   Achieving a low latency with high accuracy
        \item   Handling huge traffic without handling the switches
        \vspace{-0.15cm}
        
    \end{itemize}   &
    
    \begin{itemize}
        
        \vspace{-0.15cm}
        \item   Expensive to deploy in the existing network without changing the switch/router configuration
        \item   Specific vendor oriented
        \vspace{-0.15cm}
        
    \end{itemize}   \\  \hline

    \textit{Reversible Sketch \cite{Schweller.2007}} &
    
    \begin{itemize}
        
        \vspace{-0.15cm}
        \item   Utilizing flow header and hashing function to compact the data
        \item   Taking less memory and scalable for larger network scheme
        \vspace{-0.15cm}
        
    \end{itemize}   &
    
    \begin{itemize}
        
        \vspace{-0.15cm}
        \item   Running algorithm in each set of flows is difficult and expensive
        \item   Not supporting various measurement tasks to get the desired query
        \vspace{-0.15cm}
        
    \end{itemize}   \\  \hline

    \textit{OpenSketch \cite{MinlanYu.}} &
    
    \begin{itemize}
        
        \vspace{-0.15cm}
        \item   Having a measurement library in the control plane
        \item   Automatically configures and manage resources 
        \vspace{-0.15cm}
        
    \end{itemize}   &
    
    \begin{itemize}
        
        \vspace{-0.15cm}
        \item   lackings of generality and limited query
        \item   OpenFlow is globally accepted; deploying a new concept will be expensive
        \vspace{-0.15cm}
        
    \end{itemize}   \\  \hline

    \textit{Deltoid \cite{Cormode.2005}} &
    
    \begin{itemize}
        
        \vspace{-0.15cm}
        \item   Additional counters in each bucket are encoded in flow headers
        \item   Updating in every packet with each bucket
        \vspace{-0.15cm}
        
    \end{itemize}   &
    
    \begin{itemize}
    
        \vspace{-0.15cm}
        \item   Dramatically increasing query cost by continuous updating
        \vspace{-0.15cm}
        
    \end{itemize}   \\  \hline

    \textit{FlowRadar \cite{Li.2016b}} &
    
    \begin{itemize}
        
        \vspace{-0.15cm}
        \item   Mapping flows to counters through XOR by repeatedly XOR-ing 
        \item   Using XOR operations in network-wide recovery stage
        \vspace{-0.15cm}
        
    \end{itemize}   &
    
    \begin{itemize}
        
        \vspace{-0.15cm}
        \item    XOR-ing leads a high query cost and heavy computational overhead
        \vspace{-0.15cm}
        
    \end{itemize}   \\  \hline

    \textit{UnivMon \cite{Liu.}} &
    
    \begin{itemize}
        
        \vspace{-0.15cm}
        \item   Monitoring task to every node and data plane nodes performs the monitoring operations
        \item   Collects sketch summaries by control plane to calculate matric estimation
        \vspace{-0.15cm}
        
    \end{itemize}   &
    
    \begin{itemize}
    
        \vspace{-0.15cm}
        \item   Need to refresh components as continuously sketch collection by the controller
        \vspace{-0.15cm}
        
    \end{itemize}   \\  \hline

    \textit{SketchVisor \cite{Huang2017}} &
    
    \begin{itemize}
    
        \vspace{-0.15cm}
        \item   Utilizing two paths for flow statistics in data plane; normal path and fast path
        \item   Fast path is activated when normal path buffer is full
        \item   Control plane merges two path flows and uses traffic matrices to get network-wide recovery
        \item   Bytes of small flows can be achieved by subtracting bytes of large flows from total bytes
        \vspace{-0.15cm}
        
    \end{itemize}   &
    
    \begin{itemize}
        
        \vspace{-0.15cm}
        \item   Small flows can be missed out from the calculation as only large flows are taken into consideration
        \item   Complex mathematical model is given for retrieving small flows
        \vspace{-0.15cm}
        
    \end{itemize}   \\  \hline

    \end{tabular}
\end{table*}

\begin{itemize}

    \item   All of these traffic measurement solutions concentrated on real-time network flows collection for quick decision making.
    \item   A single framework is being used with a library of different measurement algorithms, so that network administrator can choose any combination of the algorithm to get desired traffic measurement.
    \item   Sketch-based framework is the best method though it needs a different protocol rather than OpenFlow. However, OpenFlow is now broadly acknowledged as an industry standard in data center situations and it is progressively being actualized.

\end{itemize}

%%%%%%%%%%%%%%%
\subsection{Traffic Measurement for Specificity}

In this part, we will discuss some specific measurement framework of latency measurement, bandwidth measurement and topology measurement. We already discussed about the basic idea behind these measurement activities in the early part of the paper.

\subsubsection{Latency Measurement}

A latency measurement framework \textit{SLAM} \cite{Mirkovic.2015} can quantify latency between any two-network switches along the way by sending test packets from one switch to another. In addition, it estimates latency by checking arrival packets in control plane, that is well suited to SDN engineering. Also, the accuracy of latency relies upon process time of the first and last switches along the way, which differs continuously with every occasion. The preface of estimating latency is the timestamp extricated from the packet, which implies that increasing the packet header size can be very useful for different SDN applications. There is another latency measurement framework \textit{DPTH} \cite{T.Mizrahi.2016}, that estimates latency between the two switches by sending timestamped packets from one switch to another and explicitly computes the time contrast as long as these switches have synchronized. Some of the critical characteristics of DPTH are:

\begin{itemize}
    
    \item   Mitigating color bits by adding it in the header.
    \item   Adding an extra header to all packets seems, by all accounts to be exorbitant in any. case.
    \item   Different expense in light of particular application requests.
    \item   Processing delay to measure latency accurately.
    
\end{itemize}

\begin{table*}[b!]
\small
\caption{SDN Traffic Measurement Solutions for Latency}
\label{table_specificity_latency}

    \centering
    \begin{tabular}{p{0.12\linewidth}p{0.41\linewidth}p{0.41\linewidth}} \hline
	\textbf{Reference} & \textbf{Solutions} & \textbf{Drawbacks} \\ \hline

    \textit{SLAM \cite{Mirkovic.2015}} &
    
    \begin{itemize}
    
        \vspace{-0.15cm}
        \item   Measuring the latency of any two packets by sending packets continuously from one switch to another
        \item   Defining the path which is preselected, and at the end, the controller can measure the latency by checking the control message
        \item   Quick, accurate and easily deployable without changing hardware modification
        \vspace{-0.15cm}
        
    \end{itemize}   &
    
    \begin{itemize}
        
        \vspace{-0.15cm}
        \item   Create network congestion by adding header timestamp
        \item   Need to send packets continuously
        \vspace{-0.15cm}
        
    \end{itemize}   \\  \hline

    \textit{DPTH \cite{T.Mizrahi.2016}} &
    
    \begin{itemize}
    
        \vspace{-0.15cm}
        \item   Timestamp all the packets attached with DPTH, network connected to DPTH and egress switch connected to DPTH
        \item   Using a coloring based method by adding an additional header, which can be diversified in different applications
        \vspace{-0.15cm}
        
    \end{itemize}   &
    
    \begin{itemize}
        
        \vspace{-0.15cm}
        \item   Adding additional header is costly
        \item   High processing delay
        \vspace{-0.15cm}
        
    \end{itemize}   \\  \hline

    \textit{MCPL \cite{He.2015}} &
    
    \begin{itemize}
    
        \vspace{-0.15cm}
        \item   Measuring inbound, outbound latency
        \item   After arriving the timestamp at the switch as well as host is noted down and calculate the time difference as inbound latency
        \vspace{-0.15cm}
        
    \end{itemize}   &
    
    \begin{itemize}
        
        \vspace{-0.15cm}
        \item   Need hardware updating time, installation delay
        \item   Addition and deletion are to be counted
        \vspace{-0.15cm}
        
    \end{itemize}   \\  \hline
    
    \end{tabular}
\end{table*}

Besides, \textit{MCPL} \cite{He.2015} introduced a far-reaching investigation of latency measurement, which has the efficient points such as inbound latency, outbound latency, and hardware performance inconsistency. Receiving the message, addition or subtraction, update time and installation delay is the part of outbound latency. Additionally, packet timestamp can procure the timing. We have already illustrated some latency measurement solutions which can be used for specific network application condition. Table \ref{table_specificity_latency}, shows traffic measurement solutions for latency measurement.

\subsubsection{Bandwidth Measurement}

\textit{Accessible Bandwidth (ABW)} \cite{Megyesi.2016} is the extreme packet-forwarding rate for another flow to impart the different flows. Accuracy in bandwidth measurement in SDN can quicken network administration and enhance network QoS. At the point, when controller is collecting data of the entire network, bandwidth can successfully analyze and wipe out conceivable linkage issues, which ensures that the network capacities as well \cite{Popa.2013}. 

ABW measurement is a general method for normal conditions of the network. To further improve the ABW, \textit{SOMETIME} \cite{G.Aceto.2017} is an active measurement, which is designed for wireless connectivity and applied in cellular broadband access networks, to give an estimation of ABW in SDN environment. Because of various of devices in a cellular network, the accurate ABW measurement approach is expected to represent sharing correspondence resources, and breaking constraint. With exploiting the OpenFlow protocol to focus on flows from other network traffic effortlessly, the controller is the busiest module here without affecting the much in network performance. However, presence of control message may introduce overhead, which can slow down the network.

\begin{table*}[b!]
\small
\caption{SDN Traffic Measurement Solutions for Bandwidth}
\label{table_specificity_bandwidth}

    \centering
    \begin{tabular}{p{0.15\linewidth}p{0.38\linewidth}p{0.38\linewidth}} \hline
	\textbf{Reference} & \textbf{Solutions} & \textbf{Drawbacks} \\ \hline

    \textit{ABW \cite{Megyesi.2016}} &
    
    \begin{itemize}
    
        \vspace{-0.15cm}
        \item   Measuring end to end maximum bandwidth for controlling QoS of the whole network
        \item   Solving a linkage problem and other network functions
        \vspace{-0.15cm}
        
    \end{itemize}   &
    
    \begin{itemize}
        
        \vspace{-0.15cm}
        \item   Query is dependent with the flow rate
        \item   Highly congested in the heavy traffic condition
        \vspace{-0.15cm}
        
    \end{itemize}   \\  \hline

    \textit{SOMETIME \cite{G.Aceto.2017}} &
    
        \begin{itemize}
    
        \vspace{-0.15cm}
        \item   Active method and applicable for the wireless network
        \item   Distinguishing the other network traffic flow by OpenFlow, and send it to the controller for recording
        \vspace{-0.15cm}
        
    \end{itemize}   &
    
    \begin{itemize}
        
        \vspace{-0.15cm}
        \item   Control message may lead overhead, and traffic congestion
        \item   Not generalized and straightforward framework
        \vspace{-0.15cm}
        
    \end{itemize}   \\  \hline

    \textit{MAPLE SCHEDULER \cite{R.Wang.2016}} &
    
        \begin{itemize}
    
        \vspace{-0.15cm}
        \item   Concentrating on measuring effective bandwidth to maintain QoS
        \item   Selecting edge and core part are for effective bandwidth measurement
        \item   Residual BW can be found by subtracting link capacity from effective BW
        \vspace{-0.15cm}
        
    \end{itemize}   &
    
    \begin{itemize}
        
        \vspace{-0.15cm}
        \item   Throughput is effected during measuring effective BW
        \vspace{-0.15cm}
        
    \end{itemize}   \\  \hline
    
    \end{tabular}
\end{table*}

Thus, we need ABW to be more accurate and straightforward to apply in the network applications. A way to maintain the QoS is developed by \textit{MAPLE-Scheduler} \cite{R.Wang.2016}, which is the first one measures effective bandwidth and checks routing condition. Edge and core part of the network has different effective bandwidth (EB) measurement. Also, it gives more flexibility and accuracy by deploying EB measurement in the edge part of the network. However, the disadvantage is that the accuracy cannot always be the same. Table \ref{table_specificity_bandwidth}, shows traffic measurement solutions for bandwidth measurement.

\subsubsection{Topology Discovery}

\textit{OpenFlow Discovery Protocol (OFDP)} \cite{F.Pakzad.2014} discovers network linkage hop by hop, and the entire system can be rearranged as in all switch nodes will build up TLS with the controller by handshake session in any case. At the initial stage, the controller collects the packets of full network and establish TLS session. Then the controller sends packets to the whole network with full information of data in the next stage. After all the switches getting the data, it will send packets to the nearest switches with predefined rules, and finally the controller gets the map of the full network. In addition, the controller will get the information of new topology by synchronization if any link is broken. With much efficient way to actualize the resource consumption, it troubles in high overhead.

Also, \textit{OFDP v2} \cite{S.Khan.2017} has been enhanced by restricting controller to send amount of packet numbers.Also, The switch can change the receiving packets, include MAC address and port numbers. Then these packets are sent to every single access port through switches. Although the control overhead, and CPU resources usage reduce almost a half compared to the previous version, the constraints of OFDP still exists is still not solved yet that When takes a large amount of resources while refreshing the entire network structure. It takes advantage of quality performance in high traffic.

\textit{SD-TDP} \cite{Ochoa-Aday2016} is a solution of measurement to find the topology by mitigating controllers load with breaking the entire networks into smaller parts. In the initial stage, each active node (AN) in the network keeps up deactivated state and waiting for TDP request message from the controller. The node status will be changed once a TDP-Request message is gotten with the progressive structure setting up. Then next discovery stage consists of the correspondence among father node (FN) and AN. FN has all the message and communicate with the controller, which will recover the topology structure of FN at last. When network topology changes, FN will naturally changes its state to AN. Additionally, SD-TDP relieves controllers load and enhance its performance.

\textit{NetMagic} \cite{Dai.2016} is another topology measurement platform, which can disocover actively complete network, and the controller sends test packets to all NetMagic in whole network. When NetMagic gets packets from its neighbors, it informs the controller, then the controller collects all the information from NetMagic and gets topology of the entire network. NetMagic uses hash function, which is great advantage of less using on board RAM. Besides, intradomian topology sometimes can be very complexed, which may lead overhead, and the administrator does not know what the current scenario of the network is whereas controller manages the full network. Thus, controller can not take routing decisions in intradomain topology.

\begin{table*}[b!]
\small
\caption{SDN Traffic Measurement Solutions for Topology}
\label{table_specificity_topology}

    \centering
    \begin{tabular}{p{0.15\linewidth}p{0.38\linewidth}p{0.38\linewidth}} \hline
	\textbf{Reference} & \textbf{Solutions} & \textbf{Drawbacks} \\ \hline

    \textit{OFDP \cite{F.Pakzad.2014}} &
    
    \begin{itemize}
    
        \vspace{-0.15cm}
        \item   Establishing switch nodes a connection with the controller
        \item   Sending LLDP packets by the controller and returned to discovers the full network topology
        \vspace{-0.15cm}
        
    \end{itemize}   &
    
    \begin{itemize}
        
        \vspace{-0.15cm}
        \item   Creating predefined rules of LLDP packets and update continuously make it complex to find the perfect routing paths
        \item   High resource consumption, time cost, and overhead 
        \vspace{-0.15cm}
        
    \end{itemize}   \\  \hline

    \textit{OFDP v2 \cite{S.Khan.2017}} &
    
    \begin{itemize}
    
        \vspace{-0.15cm}
        \item   Reduces sending the number of packets by the controller
        \item   Replicating these packets by adding the information in the switches, and send back to the controller
        \item   Saving 45\% overhead and 40\% CPU resources by comparing the previous version of OFDP
        \vspace{-0.15cm}
        
    \end{itemize}   &
    
    \begin{itemize}
        
        \vspace{-0.15cm}
        \item   Upgradation the full network, it takes a lot of resources
        \item   Time cost is also high
        \item   QoS control is also difficult to manage
        \vspace{-0.15cm}
        
    \end{itemize}   \\  \hline

    \textit{SD-TDP \cite{Ochoa-Aday2016}} &
    
    \begin{itemize}
    
        \vspace{-0.15cm}
        \item   Dividing the full network into two types of nodes (FN and AN)
        \item   Sending TDP messages to all AN by Controller and FN, and AN changes its status to FN
        \item   Collecting all message by FN and forward to the controller to discover the topology
        \vspace{-0.15cm}
        
    \end{itemize}   &
    
    \begin{itemize}
        
        \vspace{-0.15cm}
        \item   Performing no-action rule by FN, if the topology changed or in linkage problem, as it losts the communication with it's near AN neighbour
        \item   Problem of taking routing decision
        \vspace{-0.15cm}
        
    \end{itemize}   \\  \hline

    \textit{NetMagic \cite{Dai.2016}} &
    
    \begin{itemize}
    
        \vspace{-0.15cm}
        \item   Sending a message to NetMagics in the full network and NetMagics modify the message
        \item   Collecting all the message by the controller and stored for discovering the topology
        \item   Using NetMagic RAM to store primary data instead of using controller RAM
        \vspace{-0.15cm}
        
    \end{itemize}   &
    
    \begin{itemize}
        
        \vspace{-0.15cm}
        \item   Bringing much pressure on network scalability, if the size of the network is increased
        \item   Complex framework and difficult to design and deploy
        \vspace{-0.15cm}
        
    \end{itemize}   \\  \hline

    \textit{W. Y. Huang et al. \cite{W.Huang.2014}} &
    
    \begin{itemize}
    
        \vspace{-0.15cm}
        \item   Connecting NOX and floodlight controller by a third-party module ENVI
        \item   Saving the data structure by NOX and periodically communicates with the floodlight controller to draw an interdomain map
        \vspace{-0.15cm}
        
    \end{itemize}   &
    
    \begin{itemize}
        
        \vspace{-0.15cm}
        \item   Highly complex solutions to deploy in the congested network
        \item   Relatively lower accuracy
        \vspace{-0.15cm}
        
    \end{itemize}   \\  \hline

    \end{tabular}
\end{table*}

\textit{W. Y. Huang et. al} \cite{W.Huang.2014} utilizes an module called \textbf{ENVI} and supports a correspondence system among NOX and floodlight controller. Here, NOX will send each host interface message to ENVI and save all the data by maintaining a data structure. Also, the floodlight needs to make a correspondence with ENVI in any case and includes two data structures in floodlight controller to support the correspondence, which is an intradomain map. Along these lines, the new data structure can be put away and prepared among NOX, floodlight, and ENVI. NOX and floodlight assemble correspondence component through ENVI. Nonetheless, the enormous intricacy, low correspondence productivity, and absence of high recognizable proof accuracy of this technique still need change. Table \ref{table_specificity_topology}, shows traffic measurement solutions for topology measurement.

In a word, we analyze the existing methods as following:

\begin{itemize}

    \item   For latency measurement, control message is sent by the controller which receives the message and calculates the latency by receiving and transmitting time; so controller is the main part of whole measurement system.
    \item   For bandwidth measurement, end to end connection is used in wired and wireless network to differentiate traffic flows and send flows to the controllers.
    \item   For topology measurement, the entire network structure is partitioned into a few units, and every node in the partitions is active to process forward the control message.
    
\end{itemize}

Clearly, it is visible to find that traffic measurement in SDN is not smooth enough to perform. Researchers are still trying to solve the measurement problems to meet the minimum requirement of SDN traffic measurement solutions. Using statistic machine learning and deep learning, we can find more accurate measurement with less resource usage, low overhead, real-time and specific accurate SDN traffic measurements by using ML in a complex, hierarchical, heterogeneous next generation 5G network. In the next section, we discuss some ML based SDN traffic measurements.

\section{MACHINE LEARNING IN TRAFFIC MEASUREMENT}

In recent year, ML is popular in many fields, and this technique has been widely used by researchers to solve sophisticated problems for data science and networking. For example, it can be used in kinds of search engines like google, bing, yahoo, or web page ranking, disease analysis, network management to dig out the interests of the customers, then make decision and further analysis. Thus, ML not only can help to analyze the data but also can predict the future pattern by itself. Besides, it is a new uprising concept that it can be used in traffic measurement to learn the pattern such as routing, congestion control, QoS control, resource management, network management, fault analysis, security management, and then make future prediction of their trend. Specially in a wide area network, routing pattern is complicated to be handled by the administrator. At that point, whether the traditional networking system can be replaced by the SDN modern networking system where controllers are taking decisions for network management is the main problem. However with the highly heterogeneous network using ML, controller will be more scalable and responsive to make decisions to manage the network. A lot of survey papers \cite{Nunes.2014,Boutaba.2018,Alsheikh2014,Bkassiny2013,Buczak2016,Fadlullah2017} have applied ML to networking, and some part of them also described how we can use ML in SDN traffic measurement. Fig.~\ref{ml_measurement} gives the overview of this section.
\iffalse
\begin{figure*}
  
    \centering
    \includegraphics[width=17cm]{figure/ml_meas_ref.png}
    \caption{\label{ml_measurement}Machine learning solutions in Traffic Measurement}
\end{figure*}
\fi
\begin{figure}[t!]
\centering
\includegraphics[width=16cm]{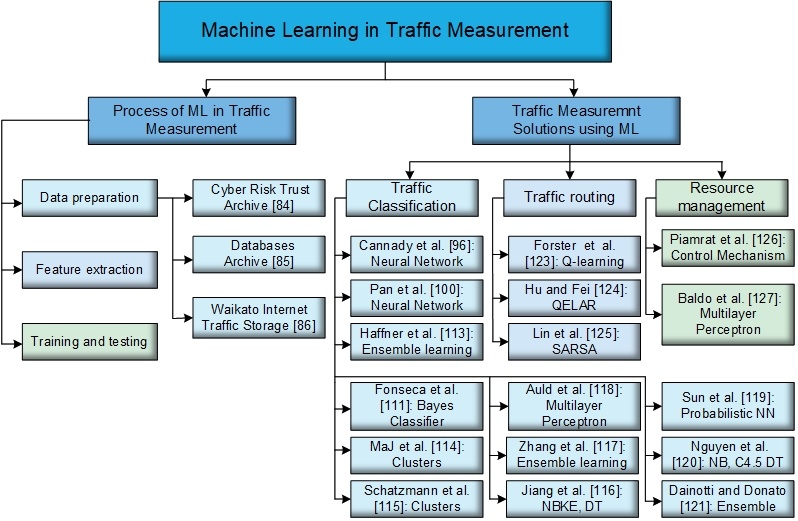}
\caption{\label{ml_measurement}Machine learning solutions in Traffic Measurement}
\end{figure}  
\subsection{Process of ML in Traffic Measurement}

\subsubsection{Data Preparation}

Before using ML in the network, we need to collect and store the packets data offline or online \cite{Wang2018} to be learned. There are various types of substantial offline data can be collected such as Cyber Risk Trust Archive \cite{cyber_risk_data}, Discovery in Databases Archive \cite{ucc_data_archive}, Waikato Internet Traffic Storage \cite{waikato_data_archive}, also the passive measurement or active measurement framework \cite{fraleigh_design_deployment_passive} can help to further store more data. Besides, this dataset will help to build a decision model to find out an accurate predictive measurement. Then the gathered data is divided into train and test dataset, and the training set is utilized to discover the perfect condition for a Neural Network (NN) as an ML model. At long last, the test set is utilized to evaluate the unprejudiced performance of the chose to demonstrate. 

\subsubsection{Feature Extraction}

During the training, we need to take the feature selection part, which is utilized to lessen dimensionality in voluminous data and to distinguish segregating features that lessen computational overhead and increment accuracy of machine learning models \cite{moore_internet_traffic_classification}. Here, feature selection and extraction can be performed utilizing different simulator, for example, NetMate \cite{yemini_netmate} ,WEKA \cite{kulkarni_weka}, Python \cite{Sisiaridis2018}, MATLAB \cite{Shakya2018}. Nonetheless, the extraction and determination procedures are constrained by the capacity of the device utilized. It is significant to deliberately choose a perfect set of features that strike a harmony between misusing connection and lessening of over-fitting for higher accuracy and lower computational overhead. Then ground truth is to be set up to relate the giving formal depiction to the classes of intrigue.

\subsubsection{Training and Testing}

After extracting features from the dataset, we train them with the machine learning methods like support vector machine (SVM), decision tree (DT), naive bayes (NB) and so on. Also, we can use these models to evaluate on the traffic measurement, and update our models each time with the online data. There are different techniques for marking datasets utilizing the features of a class. Essentially, it requires with help from deep packet inspection (DPI) \cite{Wang2016}, \cite{Zhang2013}. Once ML model has been assembled and the ground truth has been discovered, it is critical to measure the performance of the model that will anticipate, or assess outcomes. There is no real way to recognize a learning algorithm as the best model, and it is not reasonable to analyze mistake rates \cite{Jamjoom.}. The performance measurements can be utilized to gauge the distinctive parts of the model. In this paper, we discuss some ML techniques for traffic classification, prediction, routing and resource management, which can be used in SDN environment. Fig.~\ref{data_processing_ML} shows that how ML can be used in network traffic measurement.
\iffalse
\begin{figure}[t!]
    \centering
    \includegraphics[width=0.40\textwidth]{figure/data_processing_ml.png}
    \caption{\label{data_processing_ML}Machine learning process in Network Measurement}
\end{figure}
\fi
\begin{figure}[t!]
\centering
\includegraphics[width=10cm]{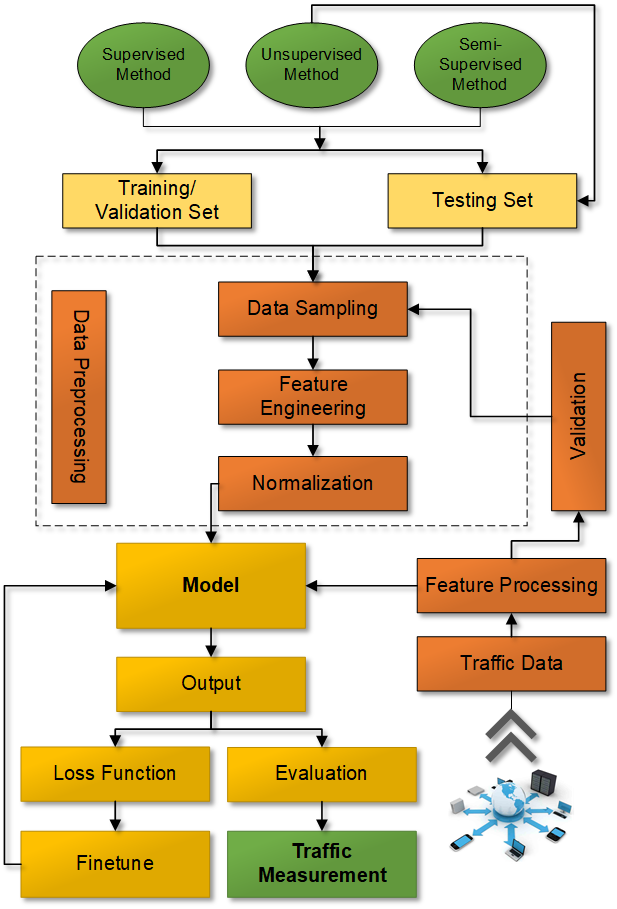}
\caption{\label{data_processing_ML}Machine learning process in Network Measurement}
\end{figure}
\subsection{Traffic Measurement Solutions using ML}

In this part, we discuss ML based SDN traffic measurement solutions and then categorize them.

\subsubsection{Traffic classification}

Today in the era of technology, where everyone is communicating through a network, and it is potential for the users to communicate or send data with the probability of being hack or damage. Thus, there are many ML based  applications that are introduced to detect misused based attack. In these applications, ML classifiers were used to distinguish the data into two groups; one is misused, and the other is anomaly based group \cite{Boutaba.2018}. ML classifier has the power to recognize the pattern of threatened attack in a large amount of data, which is coming in the network. Many studies has been conducted on the bases of ML technology in the past, to detect the misuse detection \cite{amor_intrusion_detection_systems,article,Chebrolu2005,10.1007/978-3-540-45248-5_10,Subba2016,Pan2003,Peddabachigari2007,Sangkatsanee2011,Stein:2005:DTC:1167253.1167288}. Majority work has been done in the computer (offline) like researches collected the data from the network and they process the data offline on a computer using different ML tool. Besides in the offline detection, people trained the classifiers to classify the attack data from standard data. 

\textit{Cannady et al.} \cite{article} introduced a very early virus detection system through NN which can achieve much success in saving the SDN from viruses and especially work done efficiently regarding less complexity to detect misuse data. This work has been done offline; five features were analyzed, i.e., TCP, ICMP, IP, header fields and payload. In NN, nine (9) layers were utilized to classify two categories data. The number of neurons has been determined through trial and error; the sigmoid function has been utilized as an activation function on the neurons. And It can achieve 89\% to 91\% accuracy.

\subsubsection{Traffic routing}

Besides the traffic flow classification, the traffic routing is also important in the traffic measurement to manage the traffic flows when the congestion happened. It requires challenging abilities for the ML models, such as the ability to cope and scale with complex and dynamic network topologies, the ability to learn the correlation between the selected path and the perceived QoS, and the ability to predict the consequences of routing decisions. Thus, researchers prefer applying reinforcement learning (RL) to learn the control strategy instead of human management. There are several existing papers related to the traffic routing problem using the RL methods.

In \cite{forster_feedback_routing_optimizing}, Forster et al. used a Q-learning approach in a multicast routing protocol, called FROMS (Feedback Routing for Optimizing Multiple Sinks). The goal of FROMS is to route data efficiently, in terms of hop count, from one source to many mobile sinks in a WSN by finding the optimal shared tree. And Hu and Fei \cite{hu_qelar_routing_protocol} proposed QELAR, a model-based variant of the Q-routing algorithm, to provide faster convergence, route cost reduction, and energy preservation in underwater WSNs. More recently, a centralized SARSA with a softmax policy selection algorithm has been applied by Lin et al. \cite{lin_qos_aware_routing} to achieve QoS-aware adaptive routing (QAR) in SDN.

\subsubsection{Resource management}

Besides, resource management in networking entails controlling the vital resources of the network, and manage kinds of resources to make the network stable. For programmer, it is easy to determine how to do to control the controller to take order. However, it will be more artificially if we apply machine to dig out the relationship between resources measurement and management. There are several methods using machine learning to deal with this problems.

Piamrat et al. \cite{piamrat_qoe_aware_admission} proposed an admission controlmechanism for wireless networks based on subjective quality of experience (QoE) perceived by end-users. This is in contrast to leveraging quantitative parameters, such as bandwidth, loss and latency. To do so, they first choose configuration parameters, such as codec, bandwidth, loss, delay, and jitter, along with their value ranges. Then, the authors synthetically distort a number of video samples by varying the chosen parameters. Also, Baldo et al. \cite{baldo_call_admission_control} proposed a ML based solution using MLP-NN to address the problem of user driven admission control for VoIP communications in a WLAN. In their solution, a mobile device gathers measurements on the link congestion and the service quality of past voice calls. Table \ref{table_ml_meas} shows the comparison among all the SDN traffic measurement solutions in different categories.

\begin{table*}[ht!]
\small
\caption{Summary of Machine learning based traffic measurement.}
% \vspace{-10pt}
\label{table_ml_meas}
\centering
    \begin{tabular}{p{0.25\linewidth}p{0.25\linewidth}p{0.4\linewidth}} \hline

    \textbf{Reference} & \textbf{Algorithm} & \textbf{Solution} \\ \hline
    \textit{Cannady et al.} \cite{article} & Supervised NN & To classify two categories data, they saved the network from viruses and especially work done efficiently regarding less complexity to detect misuse data \\ \hline
    \textit{Pan et al.} \cite{Pan2003} & Supervised NN and C4.5 DT & To classify two categories data, they extracted all 41 features that were processed on multilayer perceptron NN \\ \hline
    \textit{Fonseca et al.} \cite{Fonseca2005} & Supervised:Bayesian & They proposed ML approach which was used in PLC to manage congestion control  \\ \hline
    \textit{Haffner et al.} \cite{Haffner2005} & Supervised NB, AdaBoost, MaxEnt & To classify the payload and host behavior data, they apply Ensemble learning to the task \\ \hline
     \textit{MaJ et al.} \cite{Ma:2006:UMP:1177080.1177123} & Unsupervised ML & To classify the payload and host behavior data, they proved that payload based traffic classification could be done without any class label \\ \hline
     \textit{Schatzmann et al.} \cite{Schatzmann:2010:DHF:1879141.1879184} & Supervised SVM & To identify the webmail and non-webmail data, they explored the useful features and applied them to SVM \\ \hline
     \textit{Jiang et al.} \cite{Jiang2007} & Supervised NBKE & They reduced the complexity through correlation measures and symmetric measures in the domain of feature selection, which helps to make lightweight traffic classification \\ \hline
     \textit{Zhang et al.} \cite{Zhang2015} & Supervised BoF-NB & They made a hybrid model of supervised and unsupervised techniques to help classify the robust traffic  \\ \hline
     \textit{Sun et al.} \cite{Sun2007} & Supervised PNN & To make robust traffic classification, they proposed the model using probabilistic neural network model \\ \hline
     \textit{Nguyen et al.} \cite{Nguyen2012} & Supervised NB, C4.5 DT & To find the most representation features, they used NetMate \cite{yemini_netmate} to compute features \\ \hline
    \textit{Dainotti} \cite{dainotti_early_classification_traffic} \& \textit{Donato  et al.} \cite{Donato2014} &  Random Tree, RIPPER, PL, etc. & To make traffic classification, they first tested a single classifier alone, and then tested combination of classifiers \\ \hline
    \textit{Forster et al.} \cite{forster_feedback_routing_optimizing} & Q-learning algorithm & They uses a Q-learning approach in a multicast routing protocol to route data efficiently \\ \hline
    \textit{Hu et al.} \cite{hu_qelar_routing_protocol} & Q-routing algorithm & They proposed QELAR to provide faster convergence, route cost reduction, and energy preservation in underwater WSNs \\ \hline 
    \textit{Lin et al.} \cite{lin_qos_aware_routing} & SARSA algorithm & They applied  a centralized SARSA with a softmax policy selection algorithm to achieve QAR \\ \hline 
    \textit{Piamrat et al.} \cite{piamrat_qoe_aware_admission} & admission controlmechanism algorithm & They synthetically distorted a number of video samples by varying the chosen parameters \\ \hline 
    \textit{Baldo et al.} \cite{baldo_call_admission_control} & Multilayer  Perceptron & They proposed a ML-based solution to address the problem of user driven admission control for VoIP communications in a WLAN \\ \hline

    \end{tabular}
\end{table*}

\section{Future Prospects of ML in Traffic Measurement}

As the ML based methods mentioned above, the statistic machine learning algorithms can dig out the pattern of routing, congestion control, QoS and resource management to make future predictions, and then help the controller manage the network in advance. For further improvement of the performances and quality of the services, perhaps another method like deep learning or deep reinforcement learning will help to find deep features of the packets flow. In this section, we analyze the probability of different popular artificial intelligence approaches whether they can be used in traffic measurement tasks and give the future prospects of the traffic measurement in SDN.

\subsection{Statistic Machine Learning}

Statistic machine learning is the typical methods to be used in the data analysis, and we have described many exists ML based traffic measurement in the last section. Here, we will discuss some other methods whether they can be utilized in the SDN network.

\subsubsection{\textbf{SVM}}

SVM is one of the machine learning algorithm, which belongs to a generalized linear classifier, can minimize generalization error and maximize the geometric marginal. Generally in SDN traffic measurement, classifiers like NB, DT are used to recognize the features of packets data to detect attacks. However, SVM is a more useful and convenient method to make predictions, and it is robustly to kinds of cases. \cite{liu_svm_based_classification} is the example that apply SVM to the SDN to classify the application traffic, and we can transfer it to other measurement as following:

\begin{itemize}
    
    \item After receiving the packets data, SVM can be used to detect whether an attack exists.
    \item It can also spy on the resources, and predict the trend of the traffic if there will be busy in the future to maintain QoS.
    \item Also, SVM is a valuable algorithm to find the potential congestion in the network.

\end{itemize}

\subsubsection{\textbf{Decision Tree}}

Decision tree (DT) is another statistic machine learning algorithm which mainly depends on the basic dataset. It filters out the disturbing features and builds a tree to make classification with different branches. Thus, DT can be used to predict the traffic trend base on the previous knowledge which is already applied to the traffic classification \cite{Nguyen2012, yemini_netmate}. Also with online flow table, \cite{leng_flow_table_compression} utilized DT to solve the Flow Table Congestion Problem (FTCP) to guarantee the quality of service. As the same as the usage of classification or prior knowledge of congestion, Decision Tree has a wide range of cases to be applied:

\begin{itemize}
    
    \item As it generally relies on the previous data to build the decision tree, we can turn it on the resource management to help the controller control the resources.
    \item Also, it can be used to classify the received flow for better storing or understanding.

\end{itemize}

\subsubsection{\textbf{K-Nearest Neighbor}}

K-nearest neighbor (KNN) is a kind of unsupervised learning method, that has no label to use compared to the supervised methods (like SVM and DT). It can be applied to the classification or prediction tasks to divide the feature into clusters without the class knowledge. \cite{Ma:2006:UMP:1177080.1177123} utilized k-means and KNN clustering to classify the data, here we can also apply this clustering method to the task of resource management and attack detection.

\subsubsection{\textbf{Ensemble Learning}}

Ensemble learning is a strongly learnable method, that gather various of weakly learnable algorithms like NB, KNN, to improve the final prediction. Given several algorithms, they firstly dig out the features to analyse the data respectively, and then we get the final results among them by counting the highest voting. \textit{Kolomvatsos et al.} \cite{kolomvatsos_uncertainty_driven_ensemble} developed an ensemble forecasting method to provide QoS in SDN network with its own prediction rule. However, it is worse than the original machine learning algorithm. Thus in the future, we can apply ensemble learning to improve the results of tasks although it may cost time and resources:

\begin{itemize}
    
    \item First, we can gather SVM, DT, KNN or other methods to detect the attack among the flow data to make the final prediction.
    \item Next, it can also be used to classify the traffic application, and vote to get the most probably one.
    \item Also, it can learn the trend of the traffic flow, and then decides whether there will be congestion and sends instruction to the controller.

\end{itemize}

There are various types of statistic ML algorithms have not been mentioned above, and they also can be used to deal with different kinds of detection or classification tasks in SDN traffic measurement.

Despite of the statistic machine learning algorithm, DL is a new way to learn the features from the collected data to make predictions \cite{tang_neural_network_intrusion}. With the deeper layers and more neural nodes, DL can dig out deeper representations, and find relationships corresponding to the ground truth. It can also build up kinds of neural network with different functions to analyze various inputs, such as images, text or sounds. Here, we will take a look at the DL network about how we can use it into the SDN traffic measurement \cite{al2020arabic_UOB3}. 

\subsubsection{\textbf{Multilayer Perceptron}}

Multilayer Perceptron (MLP) is the basic deep learning architecture, which consists of several layers with numbers of neurons. Generally, besides input and output layer, the single hidden layer can fit kinds of linear function, and it is suitable to handle the classification tasks on text data. Moreover, MLP performs better than the statistic machine learning algorithms for its more flexibility and adaptability as well as it can also build deep architecture to find deep features. To our best knowledge that there is no paper about MLP based SDN traffic measurement, perhaps we can utilize the MLP to evaluate the text data task.

\begin{itemize}

    \item Like what the ML methods do before, we can apply MLP to the traffic classification and resources management.
    \item Also, MLP can be used to detect attack among the received data.
    \item And it can predict the congestion and traffic trend in the future.

\end{itemize}

\subsubsection{\textbf{Recurrent Neural Network}}

Recurrent neural network (RNN) is an extension of a conventional feed forward neural network (like MLP), which makes use of the sequential information. The output is depended on the previous computations in RNN and performs the same job for every single element of a sequence. It can store the memory of prior knowledge and forget the disturbing parts. \cite{tang_neural_network_intrusion} developed a gated recurrent units (GRU) approach to provide SDN network with an intrusion detection system (IDS), which can detect attacks and achieve higher performance than the ML methods. Thus, we can apply RNN with the big data analytic in SDN.

\begin{itemize}

    \item With the short memory of the studied data, RNN can take charge of the resource management with the study of the controller instructions.
    \item Long short term memory (LSTM), one of the RNN algorithms, can be taken to learn the long dependence on the data flow to classify the traffic or detect the attacks.

\end{itemize}

\subsubsection{\textbf{Convolutional Neural Network}}

Different from the deep learning methods above, convolutional neural network (CNN) is highlighted in the convolution layer, which can extract patterns from the graph. In general, CNN is mainly used to learn the classification on the image based dataset, and it can dig out the more in-depth representation with more convolution layers through the kernel filters. SDN traffic measurement in \cite{mao_network_traffic_control} applied CNN to controllers to choose the best path combination for packet forwarding in switches, where the input image is composed by the switches in time intervals. Inspired by this, we can use CNN to handle many graph based tasks.

\begin{itemize}

    \item With the collected data at each specific time, we can gather the information in the spatial dimension and build up an array data. Then we can utilize CNN to extract features from this array data to find the relationship.
    \item Also we can create 2-dimension graph with encoded features in time step, and then CNN will learn the temporal information with each column on the graph.

\end{itemize}

Thus, DL is an efficient way to make classification and prediction in traffic measurement problem, and help the network gain the prior knowledge.

\subsection{Deep Reinforcement Learning}

Different from the feature analysis with ML or DL, RL concerns more on how to lead software agents to take action time by step in an environment to maximize some notion of cumulative reward. More in details is that RL forces the agents to learn to choose the best action in each step, which will get the high scores at last. The learning process does not need any feature rather than it needs to analyze the status in each step. To further improve the performance of RL \cite{chavula_sdn_reinforcement_learning}, DRL adopts deep learning architecture \cite{liu_reinforcement_ddos_flooding} and build a deeper neural network to dig out the deep representation to find the relationship between the actions and status. Here, we will think about whether DRL can be taken to use in the traffic measurement.

\subsubsection{\textbf{Deep Q Network}}

Q-learning is a model-free algorithm using delay rewards, that interacts with the environment by perceptions and actions. It builds up a Q-table to store the reward for each action in the corresponding status. And there are many SDN measurement tasks used to deal with the congestion and multiple control problem based on Q learning \cite{zhang_qplacement_reinforcement_learning, kim_congestion_prevention_mechanism, min_dynamic_switch_migration}. However, with the rapid increase of data dimension, it is challenging to build a large Q-Table to remember the experience. Thus, Deep Q network (DQN) enables to find the low-dimensional features of high-dimensional data by crafting weights and biases in deep networks, that can replace the Q-table with the neural network. Recently \cite{qiu_blockchain_deep_qlearning} utilized DQN to consider the features of blockchain nodes and controllers jointly, and we can transfer this method into other applications:

\begin{itemize}
    
    \item Receiving attack flows, DQN can enhance the network stably by limiting the attack flows and remaining the normal communication between the users.
    \item With the congestion problem, DQN can help to find the best way to dredge flows.
    \item Although DQN can not make classification, it still can manage various of traffic to enhance the interaction.

\end{itemize}

\iffalse
\begin{figure*}[t!]
  
    \centering
    \includegraphics[width=0.9\textwidth]{figure/ml_future_research.png}
    \caption{\label{ml_future_research}Future scopes of artificial intelligence in Traffic Measurement}
\end{figure*}
\fi

\subsubsection{\textbf{Deep Deterministic Policy Gradient}}

Policy gradient is another reinforcement learning algorithm, which aims to find the best policy with each action by time step. It has more significant performance than Q-learning that it is robust to policy degradation. Besides, Deep Deterministic Policy Gradient (DDPG) improves the ability of policy gradient by applying deep neural network to dig out high dimensional features to decide the best action in each step. To our best knowledge that there is no one utilize DDPN into the SDN traffic measurement, it may improve the performance on the same tasks in DQN.

Thus, DRL is suitable to manage the SDN, and help the controller take the proper actions in different tasks. In short, the above three artificial intelligence algorithms may make progress in the traffic measurement and help to improve the network performance, which is summarized in Fig.~\ref{ml_future_research}.

\begin{figure}[t!]
\centering
\includegraphics[width=16cm]{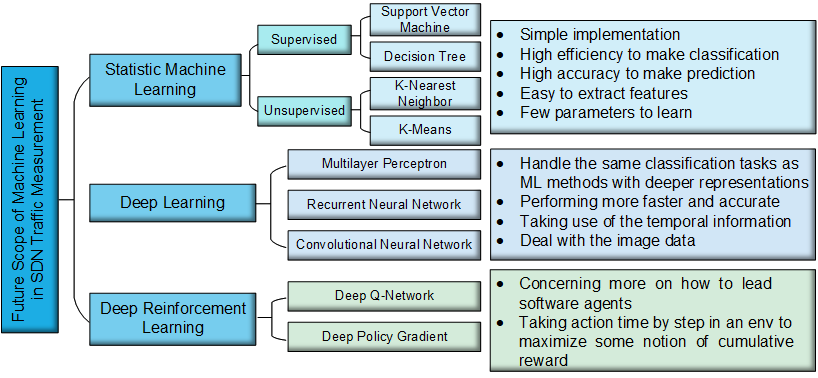}
\caption{\label{ml_future_research}Future scopes of artificial intelligence in Traffic Measurement}
\end{figure}  
\subsection{Deep Learning}

\section{FUTURE RESEARCH DIRECTION(need major revision)}

As indicated by the above foundation and technical survey, we have summarized 9-areas for further research and development. Even though SDN has profits by the network security viewpoint, despite everything, a few detectable feeble focuses have been located. The following parts ought to be the future research issues, which are summarized in Fig.~\ref{future_research_direction}.
\iffalse
\begin{figure*}
  
    \centering
    \includegraphics[width=1\textwidth]{figure/future_research_scope.png}
    \caption{\label{future_research_direction}Future Research in nine areas}
\end{figure*}
\fi

\begin{figure}[t!]
\centering
\includegraphics[width=16cm]{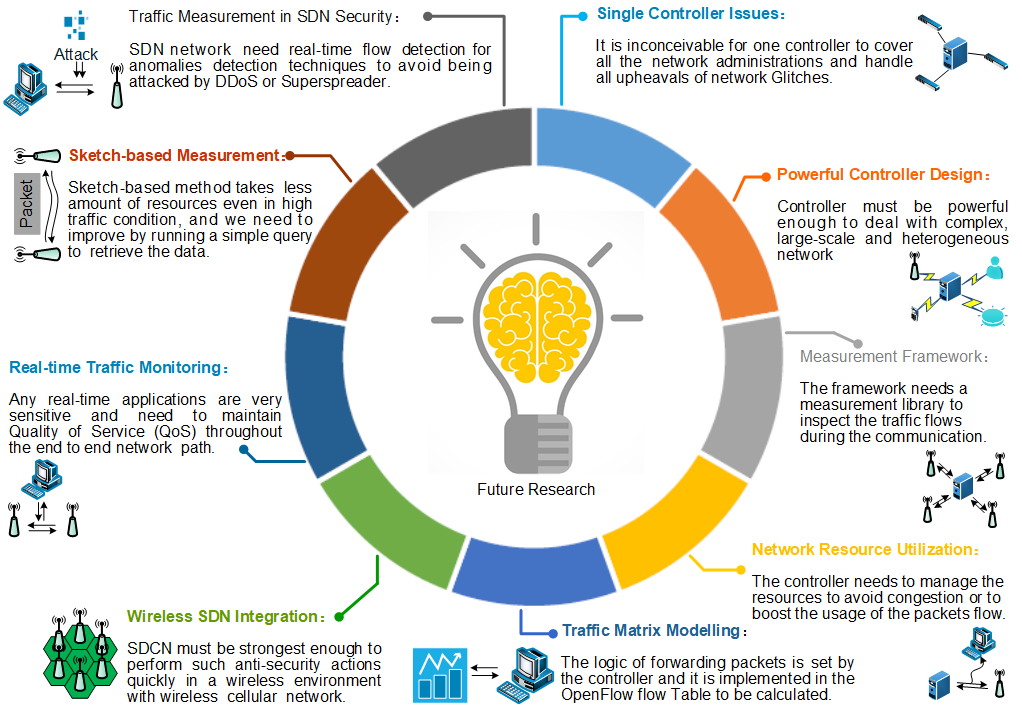}
\caption{\label{future_research_direction}Future Research in nine areas}
\end{figure}
\subsection{Single Controller Issues}

First of all, because of the exorbitant size of the network, it is inconceivable for one controller to cover all the network administrations and handle all upheavals of network glitches \cite{zhang_resource_saving_replication}. Thus, if a controller can not control the control plane or if the controller breaks down, routers/switches will be disconnected from the network. This comes to the coordination among controllers, which ought to be a fascinating and important research area. Once a network contains different territories, the trouble of identifying network security by SDN traffic measurement solutions will develop exponentially. Along these lines, analysts should figure out new sharp methodologies appropriate for dealing with this issue.

\begin{itemize}

    \item Single controller can introduce overhead and congestion in high traffic load. However, multi-controller can solve this problem by improving fault tolerance of the network. If one controller loses the connection with the network, the backup controller needs to take over the network without compromising the propagation delay. 
    \item Multi-controllers need proper co-ordination with each other. We need specific metric to formulate survivability and resource usage policy to recover the network failure under high traffic load.
    \item Controller must consider about scalability of the heterogenous network. Thus, we need to create important controller and backup controller location for considering the network failure recovery.
    
\end{itemize}
%%%%%%%%%%%%
\subsection{Powerful Controller Design}

As all switches and routers are simple forwarding devices in SDN environment, the controller takes action for all network functions. Thus, SDN controller must be powerful in sense of hardware and software because of its software-centric functionality. SDN controller is the key element of the network as all the network devices are logically centralized by a controller, and different network functions, control applications, virtualizations also are deployed in SDN environment through the controller. Thus, controller must be strong enough to tackle the load of heterogeneous network \cite{adami_novel_sdn_controller}.

\begin{itemize}
    
    \item Controller needs to support various types of network functions, control applications, virtualization functions. These types of applications may help the network controllable and scalable.
    \item Controller capacity is one of the prime concerns for hardware design, so that it can support the whole network in high traffic load. Also, QoS control, traffic recovery, traffic load balancing, and resource usage policy need to be deployed by controller to the network. In such case, the capacity and performance of controller plays an crucial role.
    \item ML and DL can be used for packet flow control in a controller, and controller work load can be distributed to some other switches and routers.
    
\end{itemize}

\subsection{Measurement Framework}

A comprehensive framework for SDCN security is necessary to establish for overcoming the challenges, and SDN traffic measurement plays a important role in this framework. As we discussed before, some recently proposed frameworks are suggested to use the library for different measurement function. It will be fascinating if these functions can store in virtual server and imply on the network with the help of controller. As traffic management is the next stage of the traffic measurement, without a proper framework, measurement activities can make network dull and introduce overhead \cite{shu_traffic_measurement_management}.

\begin{itemize}
    
    \item A single framework can be used for all measurement tasks. As we discussed, traffic management is going to be very difficult without proper monitoring by traffic measurement such as general framework, traffic prediction framework, parameter measurement framework.
    \item Library in the framework can be used for different combination of the algorithms to find out the required measurement jobs. Some pre-selected model algorithms will be saved in the library and later on, network administrator can make some useful mix-up procedure from the library to get perfect measurement.
    \item Different measurement tasks with the combination of algorithms can be saved in cloud server, which is very effective in a virtual environment. Even if the administrator are not with the physical environment, measurement activities are going on ceaselessly from a virtual environment to monitor the network.
    
\end{itemize}

\subsection{Network Resource Utilization}

SDN traffic measurement has a flexible flow measurement with different granularities and needs to satisfy several types of applications. Designing a large scale network is complex because it needs traffic forecasting, prediction, and estimation. Thus as a result, because of network resources are not used properly sometimes, link utilization becomes non-optimal\cite{T_SDN_load}. In a scenario like a little bandwidth is being provisioned without thinking of future administrations that may perhaps require more bandwidth. Afterward, when we need to arrangement high-bandwidth, unavoidably less optimum ways will be utilized for them. In such cases, network devices are expected to give scalable and dynamic measurement plots that would educate network administrators and enable them to boost their resource usage. Considering the overhead issue, it requires looking into in deep learning in sampling systems to find out about flows, assessing their size, proficient capacity, and algorithmic methodologies for the controller. Here, we can use network resource utilization as the following:

\begin{itemize}
    
    \item ML and DL approach can be used for proper network resource distribution, which can be taken a powerful sampling technique. Also, optimization techniques can be used for bandwidth provisioning.
    \item Network administrator needs the power to manipulate the network resources by considering the requirement of scalability and granularity. Both CPU and memory based approaches can help load balancing, reliable network connection and higher resource utilization \cite{hamed_novel_resource_utilization}.
    \item It needs efficient storage for flexible flow measurement and different simple low overhead algorithmic approaches for flow accounting. Dedicated applications for measurement can manage bandwidth properly for network resource utilization.
    
\end{itemize}

\subsection{Traffic Matrix Modelling}

Traffic matrix estimation and modeling are always a technical challenge for researchers. Recently, researchers are struggled to find out a way of getting an error-free model that can help to retrieve the data as well. OpenFlow has a special feature of flow table counters that allow other mechanisms to estimate traffic matrix. Thus, OpenFlow opens the door of research in this mathematical concepts. The logic of forwarding packets is set by the controller for implementation in the OpenFlow flow table, which is the key to traffic matrix calculation. Therefore, we need a statistical method to get traffic matrices which can make SDN more scalable \cite{li_estimating_sdn_traffic}.

\begin{itemize}
    
    \item Traditional traffic matrix estimation has high errors to adapt the network characteristics in varying environments. Thus, we need to use more accurate statistical methods to estimate the traffic matrix.
    \item A simple and scalable algorithm for greater accuracy in high traffic congestion can make network more robust. The CPU usage will be lower and the response time will be fast enough to recover almost all the traffic accurately. 
    \item ML and DL can be used as probabilistic approach in network wide recovery. These two models are robust to get prediction with high accuracy.
    
\end{itemize}

\subsection{Wireless SDN Integration}

We have discussed so far about SDCN, which can be a significant part of mobile network. SDN integration in wireless cellular network is always challenging to deploy because it is really very difficult to control in a wireless environment \cite{haque_wireless_sdn_survey}. It creates more vulnerabilities in the network. SDCN must be strongest enough to perform such anti-security actions quickly in a wireless environment. Different types of wireless networks such as cellular, mesh, sensor and home network are used nowadays and traffic measurement in this area is inevitable.

\begin{itemize}

    \item We can use network functions in a cloud as network functions virtualization instead of directly using in SDN control plane. An efficient virtualization strategy can control dynamic radio access with an interface.
    \item Network administrator needs to configure the network remotely to fix out the network issues. 
    \item The core network of cellular network can be controlled fully by high programmable controller. Programming capability can be used to control user data, authentication and application plane. Thus, open source idea can change the way of designing requirement of cellular network.
    
\end{itemize}

\subsection{Real-time Traffic Monitoring}

The real-time applications are very sensitive and need to maintain QoS throughout the end to end network path. Video chatting, wireless mobile applications, online learning applications are some of the examples of its applications, which need real-time monitoring and traffic measurement \cite{tse_sdn_enabled_core}. Though SDN provides a real-time centralized controller, the network operator must deal with scalability issue, and it is quite difficult to maintained fine-grained measurement \cite{zhou_real_time_sdn}.Network administrator also needs to manage this by constructing APIs that run on the controller to take actions automatically by measuring traffic data \cite{T_SDN_MGM}. Researchers may focus on designing such kind of APIs in the future.

\begin{itemize}
    
    \item As we need to use multi-controller in a huge network, synchronization of all the controllers' real time traffic monitoring is a prime concern to overcome the network congestion.
    \item We can use statistical prediction method in real time traffic measurement. Machine learning model can help to forecast the traffic flow statistics in a high traffic load, which helps the controller to make early decision.
    \item We can take dedicated applications in real time QoS and SLA monitoring. Some networks and control applications can also make the network more secured and scalable.
    
\end{itemize}

\subsection{Sketch-based Measurement}

We have discussed different types of measurement solutions in the sections before. However, we find that sketch-based solution is one of the best solutions in SDN traffic measurement, which takes less amount of resources even in high traffic condition. However, using less resource \cite{su_costa_cross_layer, kaplan_development_sketch_based} causes some problem in network recovery stage, and it is quite difficult to retrieve all data in control plane. Therefore, it is possible to make a new design idea in control plane so that it can retrieve the data by running a simple query.

\begin{itemize}
    
    \item As we discussed, sketch based measurement solutions have a problem in network recovery stage. Therefore, we can use ML method to predict packet flow statistics and then we can get high accuracy in measurement.
    \item This framework can be used in wide area network for flow monitoring and network management. Sketch based measurement needs to be deployed in a complex and heterogeneous network, so that it can perform in real life scenerio.
    \item We need to deploy sketch visor with other framework with high packet forwarding performance to give more efficiency in SDN measurement.
    
\end{itemize}

\subsection{Traffic Measurement in SDN Security}

SDN has been widely utilized because of its simplicity to control over heterogeneous and complex network. At the same time, different institutions such as bank, business organizations, financial organizations are also starting to use SDN into their core network. We all know that, this sophisticated network is connected with different servers with valuable, crucial and private information, which is always in a risk of data theft. Currently, there are a few SDN measurement approaches \cite{monshizadeh_adaptive_detection_prevention, rebecchi_traffic_monitoring_ddos} available to ensure the risk assurance, which we have already discussed before. Different types of attacks such as DDoS, superspreader can take over the control of the network easily by pushing down the controller. Thus, we need real-time flow detection, and anomalies detection techniques to recover the affected controller.

\begin{itemize}
    
    \item We need to develop applications to collect traffic flow statistics data and analyze packets that are not protocol conformant.
    \item It is necessary to develop real time anomalies detection techniques at the flow level because flow level statistics is crucial for sdn security.
    \item ML and DL is going to be the most essential part for predicting flow level and take the decision with the controller before being attacked by the attacker.
    
\end{itemize}

\section{Conclusion}
In this paper, we give a review of SDN and SDCN structures and OpenFlow network, then discuss about the traffic measurement task to identify network security, resources management, quality of service and so on. To compare the existing works, we surveyed the state of the art of the methods for measuring the SDN traffic and analysis their advantages and disadvantages. Furthermore, we break down security challenges of SDN and SDN traffic measurement requirements to get highly secured SDCN, where network administration advancements must be a robust and proficient approach to manage those difficulties. Besides, we proposed to utilize ML algorithms to improve the measurement performance, and we also investigate the existing ML based methods in SDN network. At last, we make a future prospects about how we can apply artificial intelligence into the SDN traffic measurement, and analyze the developing trend of the SDN, SDCN network. As a result, traffic measurement in SDN is still challenging and few difficulties require to be handled. As SDCN will be next-generation mobile network, we have to consider the many issues and perhaps the ML may plays an important role in it.

%Bibliography
\bibliographystyle{unsrt}  
\bibliography{references}

\begin{thebibliography}{100}

\bibitem{sun2017sdpa}
Chen Sun, Jun Bi, Haoxian Chen, Hongxin Hu, Zhilong Zheng, Shuyong Zhu, and
  Chenghui Wu.
\newblock Sdpa: Toward a stateful data plane in software-defined networking.
\newblock {\em IEEE/ACM Transactions on Networking (TON)}, 25(6):3294--3308,
  2017.

\bibitem{SoftwareDefinedNetworks:TheNewNormofNetworks.}
{Software Defined Networks: The New Norm of Networks}.
\newblock {Open Networking Foundation, White paper, 2012}.

\bibitem{opendaylight}
{https://www.linuxfoundation.org/projects/case-studies/opendaylight/}.
\newblock {Open DayLight}, 2016.

\bibitem{majeed2021spike_UOB2}
Anwar~Dhyaa Majeed and Nadia Adnan~Shiltagh Al-Jamali.
\newblock Spike neural network as a controller in sdn network.
\newblock {\em Journal of Engineering}, 27(9):64--77, 2021.

\bibitem{Erickson.}
David Erickson.
\newblock {The beacon openflow controller}.
\newblock In {\em ACM SIGCOMM workshop on Hot topics in software defined
  networking}, page~13, 2013.

\bibitem{Casado2007}
Martin Casado, Michael~J Freedman, Justin Pettit, Jianying Luo, Nick McKeown,
  and Scott Shenker.
\newblock {Ethane: taking control of the enterprise}.
\newblock In {\em ACM Sigcomm}, pages 1--12, 2007.

\bibitem{blenk_network_virtualization}
Andreas Blenk, Arsany Basta, Martin Reisslein, and Wolfgang Kellerer.
\newblock {Survey on network virtualization hypervisors for software defined
  networking}.
\newblock {\em IEEE Communications Surveys and Tutorials}, 18(1):655--685,
  2016.

\bibitem{Heller2010}
Brandon Heller, Srinivasan Seetharaman, Priya Mahadevan, Yiannis Yiakoumis,
  Puneet Sharma, Sujata Banerjee, and Nick McKeown.
\newblock {ElasticTree: Saving Energy in Data Center Networks.}
\newblock {\em NSDI}, 10:249--264, 2010.

\bibitem{T}
Tariq~Emad Ali, Ameer~Hussein Morad, and Mohammed~A. Abdala.
\newblock Efficient private cloud resources platform.
\newblock In {\em 2021 International Conference on Electrical, Communication,
  and Computer Engineering (ICECCE)}, pages 1--6, 2021.

\bibitem{amin2018hybrid}
Rashid Amin, Martin Reisslein, and Nadir Shah.
\newblock Hybrid sdn networks: A survey of existing approaches.
\newblock {\em IEEE Communications Surveys \& Tutorials}, 20(4):3259--3306,
  2018.

\bibitem{akyildiz2016research}
Ian~F Akyildiz, Ahyoung Lee, Pu~Wang, Min Luo, and Wu~Chou.
\newblock Research challenges for traffic engineering in software defined
  networks.
\newblock {\em IEEE Network}, 30(3):52--58, 2016.

\bibitem{Newman.1998}
Peter Newman, Greg Minshall, and Thomas~L. Lyon.
\newblock {IP switching - ATM under IP}.
\newblock {\em IEEE/ACM Transactions on Networking}, 6(2):117--129, 1998.

\bibitem{N.Gudeetal..2008}
Natasha Gude, Teemu Koponen, Justin Pettit, Ben Pfaff, Martin Casado, Nick
  McKeown, and Scott Shenker.
\newblock {NOX: towards an operating system for networks}.
\newblock {\em SIGCOMM Computer Communication Review}, 38(3):105--110, 2008.

\bibitem{Jamjoom.}
Hani Jamjoom, Dan Williams, and Upendra Sharma.
\newblock {Don't call them middleboxes, call them middlepipes}.
\newblock In {\em Proceedings of the third workshop on Hot topics in software
  defined networking}, pages 19--24, 2014.

\bibitem{McKeown.2008}
Nick McKeown, Tom Anderson, Hari Balakrishnan, Guru Parulkar, Larry Peterson,
  Jennifer Rexford, Scott Shenker, and Jonathan Turner.
\newblock {OpenFlow: Enabling innovation in campus networks}.
\newblock {\em ACM SIGCOMM Computer Communication Review}, 38(2), 2008.

\bibitem{A.Doriaetal.2010}
L.~Yang, R.~Dantu, T.~Anderson, and R.~Gopal.
\newblock {Forwarding and Control Element Separation (ForCES) Protocol
  Specification}, 2010.

\bibitem{Song2014}
Haoyu Song, Jun Gong, Hongfei Chen, and Justin Dustzadeh.
\newblock {Unified POF Programming for Diversified SDN Data Plane}.
\newblock {\em ICNS}, 2015.

\bibitem{Nunes.2014}
Bruno Astuto~A. Nunes, Marc Mendonca, Xuan~Nam Nguyen, Katia Obraczka, and
  Thierry Turletti.
\newblock {A survey of software-defined networking: Past, present, and future
  of programmable networks}.
\newblock {\em IEEE Communications Surveys and Tutorials}, 16(3):1617--1634,
  2014.

\bibitem{Lara.2014}
Adrian Lara, Anisha Kolasani, and Byrav Ramamurthy.
\newblock {Network innovation using open flow: A survey}.
\newblock {\em IEEE Communications Surveys and Tutorials}, 16(1):493--512,
  2014.

\bibitem{S.T.Ali.2015}
Syed~Taha Ali, Vijay Sivaraman, Adam Radford, and Sanjay Jha.
\newblock {A Survey of Securing Networks Using Software Defined Networking}.
\newblock {\em IEEE Transactions on Reliability}, 64(3):1086--1097, 2015.

\bibitem{Jarraya.2014}
Yosr Jarraya, Taous Madi, and Mourad Debbabi.
\newblock {A survey and a layered taxonomy of software-defined networking}.
\newblock {\em IEEE Communications Surveys and Tutorials}, 16(4):1955--1980,
  2014.

\bibitem{GSMAssociation.2015}
Christian Olsen.
\newblock {The Mobile Economy.}, 2008.

\bibitem{Ogul2013}
Murat O{\u{g}}ul and Selçuk Baktır.
\newblock {Practical Attacks on Mobile Cellular Networks and Possible
  Countermeasures}.
\newblock {\em OALib Journal of Future Internet}, 5(4):474--489, 2013.

\bibitem{S.Mavoungou.2016}
Silvere Mavoungou, Georges Kaddoum, Mostafa Taha, and Georges Matar.
\newblock {Survey on threats and attacks on mobile networks}.
\newblock {\em IEEE Access}, 4:4543--4572, 2016.

\bibitem{haque_wireless_sdn_survey}
I~T Haque and N~Abu-Ghazaleh.
\newblock {Wireless Software Defined Networking: A Survey and Taxonomy}.
\newblock {\em IEEE Communications Surveys {\&} Tutorials}, 18(4):2713--2737,
  2016.

\bibitem{amin2016auto}
Rashid Amin, Nadir Shah, Babar Shah, and Omar Alfandi.
\newblock Auto-configuration of acl policy in case of topology change in hybrid
  sdn.
\newblock {\em IEEE Access}, 4:9437--9450, 2016.

\bibitem{sung2011towards}
Yu-Wei~Eric Sung, Xin Sun, Sanjay~G Rao, Geoffrey~G Xie, and David~A Maltz.
\newblock Towards systematic design of enterprise networks.
\newblock {\em IEEE/ACM Transactions on Networking (TON)}, 19(3):695--708,
  2011.

\bibitem{gharsallahsdn}
Amina Gharsallah, Faouzi Zarai, and Mahmoud Neji.
\newblock Sdn/nfv-based handover management approach for ultradense 5g mobile
  networks.
\newblock {\em International Journal of Communication Systems}, page e3831,
  2019.

\bibitem{nguyen2017sdn}
Van-Giang Nguyen, Anna Brunstrom, Karl-Johan Grinnemo, and Javid Taheri.
\newblock Sdn/nfv-based mobile packet core network architectures: A survey.
\newblock {\em IEEE Communications Surveys \& Tutorials}, 19(3):1567--1602,
  2017.

\bibitem{Li.2016}
Wenjuan Li, Weizhi Meng, and Lam~For Kwok.
\newblock {A survey on OpenFlow-based Software Defined Networks: Security
  challenges and countermeasures}.
\newblock {\em Journal of Network and Computer Applications}, 68:126--139,
  2016.

\bibitem{Tang.2018}
Jiawei Tang, Anfeng Liu, Ming Zhao, and Tian Wang.
\newblock {An Aggregate Signature Based Trust Routing for Data Gathering in
  Sensor Networks}.
\newblock {\em Security and Communication Networks}, pages 1--30, 2018.

\bibitem{ChuYuHunag.2010}
Chu Yu~Hunag, Tseng Min~Chi, Chen Yao~Ting, Chou Yu~Chieh, and Chen Yan~Ren.
\newblock {A novel design for future on-demand service and security}.
\newblock In {\em International Conference on Communication Technology
  Proceedings, ICCT}, pages 385--388, 2010.

\bibitem{Tso.2013b}
Fung~Po Tso and Dimitrios~P. Pezaros.
\newblock {Improving data center network utilization using near-optimal traffic
  engineering}.
\newblock {\em IEEE Transactions on Parallel and Distributed Systems},
  24(6):1139--1148, 2013.

\bibitem{Javadtalab.2015}
Abbas Javadtalab, Mehdi Semsarzadeh, Aziz Khanchi, Shervin Shirmohammadi, and
  Abdulsalam Yassine.
\newblock {Continuous one-way detection of available bandwidth changes for
  video streaming over best-effort networks}.
\newblock {\em IEEE Transactions on Instrumentation and Measurement},
  64(1):190--203, 2015.

\bibitem{Kim.2013}
Hyojoon Kim and Nick Feamster.
\newblock {Improving network management with software defined networking}.
\newblock {\em IEEE Communications Magazine}, 51(2):114--119, 2013.

\bibitem{V.MohanY.R.J.ReddyandK.Kaplan.2011}
Venkat Mohan, Y~R~Janardhan Reddy, and K~Kalpana.
\newblock {Active and Passive Network Measurements : A Survey}.
\newblock {\em Computer Science and Information Technologies}, 2(4):1372--1385,
  2011.

\bibitem{Sezer.2013}
Sakir Sezer, Sandra Scott-Hayward, Pushpinder Chouhan, Barbara Fraser, David
  Lake, Jim Finnegan, Niel Viljoen, Marc Miller, and Navneet Rao.
\newblock {Are we ready for SDN? Implementation challenges for software-defined
  networks}.
\newblock {\em IEEE Communications Magazine}, 51(7):36--43, 2013.

\bibitem{M.JarchelT.ZinnerT.HohnandP.TranGia.2013}
Michael Jarschel, Thomas Zinner, Thomas Hohn, and Phuoc Tran-Gia.
\newblock {On the accuracy of leveraging SDN for passive network measurements}.
\newblock In {\em Australasian Telecommunication Networks and Applications
  Conference (ATNAC)}, pages 41--46, 2013.

\bibitem{ali2020traffic_Ta_Traffic}
Tariq~Emad Ali, Ameer~Hussein Morad, and Mohammed~A Abdala.
\newblock Traffic management inside software-defined data centre networking.
\newblock {\em Bulletin of Electrical Engineering and Informatics},
  9(5):2045--2054, 2020.

\bibitem{MinlanYu.}
Minlan Yu, Lavanya Jose, and R~Miao.
\newblock {Software defined traffic measurement with opensketch}.
\newblock {\em 10th USENIX Symposium on Networked Systems}, pages 29--42, 2013.

\bibitem{Liu.}
Zaoxing Liu, Antonis Manousis, Gregory Vorsanger, Vyas Sekar, and Vladimir
  Braverman.
\newblock {One Sketch to Rule Them All: Rethinking Network Flow Monitoring with
  UnivMon}.
\newblock In {\em Proceedings of the conference on ACM SIGCOMM}, pages
  101--114, 2016.

\bibitem{Moshref.2014}
Masoud Moshref, Minlan Yu, Ramesh Govindan, and Amin Vahdat.
\newblock {Dream: Dynamic Resource Allocation for Software-defined
  Measurement}.
\newblock In {\em ACM conference on SIGCOMM}, pages 419--430, 2014.

\bibitem{Moshref.2015}
Masoud Moshref, Minlan Yu, Ramesh Govindan, and Amin Vahdat.
\newblock {Scream: Sketch Resource Allocation for Software-defined
  Measurement}.
\newblock In {\em Proceedings of the 11th ACM Conference on Emerging Networking
  Experiments and Technologies}, pages 1--13, 2015.

\bibitem{Moshref.2016}
Masoud Moshref, Minlan Yu, Ramesh Govindan, and Amin Vahdat.
\newblock {Trumpet: Timely and Precise Triggers in Data Centers}.
\newblock In {\em Proceedings of the conference on ACM SIGCOMM}, pages
  129--143, 2016.

\bibitem{jose_online_measurement_traffic}
Lavanya Jose, Minlan Yu, and Jennifer Rexford.
\newblock {Online measurement of large traffic aggregates on commodity
  switches}.
\newblock In {\em Proc. of the USENIX HotICE workshop}, page~13, 2011.

\bibitem{Bakshi2013}
Kapil Bakshi.
\newblock {Considerations for Software Defined Networking (SDN): Approaches and
  use cases}.
\newblock In {\em IEEE Aerospace Conference Proceedings}, pages 1--9, 2013.

\bibitem{Alhanani2014}
Rafat~Ahmed Alhanani and Jaafar Abouchabaka.
\newblock {An overview of different techniques and algorithms for network
  topology discovery}.
\newblock In {\em Second World Conference on Complex Systems (WCCS)}, pages
  530--535, 2014.

\bibitem{nehra2018tilak}
Ajay Nehra, Meenakshi Tripathi, Manoj~Singh Gaur, Ramesh~Babu Battula, and
  Chhagan Lal.
\newblock Tilak: A token-based prevention approach for topology discovery
  threats in sdn.
\newblock {\em International Journal of Communication Systems}, page e3781,
  2018.

\bibitem{wang_bandwidth_allocation_strategy}
Zhiwen Wang and Hongtao Sun.
\newblock {Bandwidth Allocation Strategy of Networked Control System based on
  Multirate Sampling Method}.
\newblock In {\em International Journal of Digital Content Technology and its
  Applications}, volume~6, pages 651--659, 2012.

\bibitem{paul2016enhanced}
Anup~Kumar Paul, Atsuo Tachibana, and Teruyuki Hasegawa.
\newblock An enhanced available bandwidth estimation technique for an
  end-to-end network path.
\newblock {\em IEEE Transactions on Network and Service Management},
  13(4):768--781, 2016.

\bibitem{megyesi2017challenges}
P{\'e}ter Megyesi, Alessio Botta, Giuseppe Aceto, Antonio Pescap{\'e}, and
  S{\'a}ndor Moln{\'a}r.
\newblock Challenges and solution for measuring available bandwidth in software
  defined networks.
\newblock {\em Computer Communications}, 99:48--61, 2017.

\bibitem{Zhang2010}
Ningbo Zhang, Fengyu Wang, Bin Gong, and Liangxiong Li.
\newblock {Identifying heavy-hitter flows fast and accurately}.
\newblock In {\em 2nd International Conference on Future Computer and
  Communication}, volume~3, pages 3--26, 2010.

\bibitem{Nagpal2015}
B~Nagpal, P~Sharma, N~Chauhan, and A~Panesar.
\newblock {DDoS tools: Classification, analysis and comparison}.
\newblock In {\em 2nd International Conference on Computing for Sustainable
  Global Development (INDIACom)}, pages 342--346, 2015.

\bibitem{Kamiyama2007}
Noriaki Kamiyama, Tatsuya Mori, and Ryoichi Kawahara.
\newblock {Simple and adaptive identification of superspreaders by flow
  sampling}.
\newblock In {\em Proceedings - IEEE INFOCOM}, pages 2481--2485, 2007.

\bibitem{SHI2014}
H~SHI, R~V Prasad, E~Onur, and I~G M~M Niemegeers.
\newblock {Fairness in Wireless Networks:Issues, Measures and Challenges}.
\newblock {\em IEEE Communications Surveys {\&} Tutorials}, 16(1):5--24, 2014.

\bibitem{salman2020link_UOB}
Mustafa~Ismael Salman et~al.
\newblock Link failure recovery for a large-scale video surveillance system
  using a software-defined network.
\newblock {\em Journal of Engineering}, 26(1):104--120, 2020.

\bibitem{T_SDN_DC}
Tariq~Emad Ali, Ameer~Hussein Morad, and Mohammed~A Abdala.
\newblock Sdn implementation in data center network.
\newblock {\em Journal of Communications}, 14(3):223--228, 2019.

\bibitem{Malboubi.2014}
Mehdi Malboubi, Liyuan Wang, Chen~Nee Chuah, and Puneet Sharma.
\newblock {Intelligent SDN based traffic (de)Aggregation and Measurement
  Paradigm (iSTAMP)}.
\newblock In {\em IEEE INFOCOM}, pages 934--942, 2014.

\bibitem{vanAdrichem.2014}
Niels~L.M. Van~Adrichem, Christian Doerr, and Fernando~A. Kuipers.
\newblock {OpenNetMon: Network monitoring in OpenFlow software-defined
  networks}.
\newblock In {\em IEEE/IFIP Network Operations and Management Symposium:
  Management in a Software Defined World}, 2014.

\bibitem{Tootoonchian.2010}
Amin Tootoonchian, Monia Ghobadi, and Yashar Ganjali.
\newblock {OpenTM: Traffic Matrix Estimator for OpenFlow Networks}.
\newblock In {\em Passive and Active Measurement: 11th International
  Conference, Zurich, Switzerland}, pages 201--210, 2010.

\bibitem{Moshref.2013}
Masoud Moshref, Minlan Yu, and Ramesh Govindan.
\newblock {Resource/accuracy tradeoffs in software-defined measurement}.
\newblock In {\em Proceedings of the second ACM SIGCOMM workshop on Hot topics
  in software defined networking}, 2013.

\bibitem{Zhang.2013}
Ying Zhang.
\newblock {An adaptive flow counting method for anomaly detection in SDN}.
\newblock In {\em Proceedings of the ninth ACM conference on Emerging
  networking experiments and technologies}, pages 25--30, 2013.

\bibitem{Chowdhury.2014}
Shihabur~Rahman Chowdhury, Md.~Faizul Bari, Reaz Ahmed, and Raouf Boutaba.
\newblock {PayLess: A low cost network monitoring framework for Software
  Defined Networks}.
\newblock In {\em IEEE Network Operations and Management Symposium (NOMS)},
  pages 1--9, 2014.

\bibitem{Dusi.2014}
Maurizio Dusi, Roberto Bifulco, Francesco Gringoli, and Fabian Schneider.
\newblock {Reactive logic in software-defined networking: Measuring flow-table
  requirements}.
\newblock In {\em 10th International Wireless Communications and Mobile
  Computing Conference}, pages 340--345, 2014.

\bibitem{Tso.2013}
Fung~Po Tso and Dimitrios~P. Pezaros.
\newblock {Baatdaat: Measurement-based flow scheduling for cloud data centers}.
\newblock In {\em Proceedings - International Symposium on Computers and
  Communications}, pages 765--770, 2013.

\bibitem{Sun.2015}
Peng Sun, Minlan Yu, Michael~J. Freedman, Jennifer Rexford, and David Walker.
\newblock {HONE: Joint Host-Network Traffic Management in Software-Defined
  Networks}.
\newblock {\em Journal of Network and Systems Management}, 23(2):374--399,
  2015.

\bibitem{Rasley.2014}
Jeff Rasley, Brent Stephens, Colin Dixon, Eric Rozner, Wes Felter, Kanak
  Agarwal, John Carter, and Rodrigo Fonseca.
\newblock {Planck: Millisecond-scale Monitoring and Control for Commodity
  Networks}.
\newblock In {\em ACM conference on SIGCOMM}, pages 407--418, 2014.

\bibitem{Suh.2014}
Junho Suh, Ted~Taekyoung Kwon, Colin Dixon, Wes Felter, and John Carter.
\newblock {OpenSample: A low-latency, sampling-based measurement platform for
  commodity SDN}.
\newblock In {\em IEEE International Conference on Distributed Computing
  Systems}, pages 228--237, 2014.

\bibitem{MakingtheNetworkVisible.}
{SFlow}.
\newblock {www.sflow.org}.

\bibitem{Schweller.2007}
Robert Schweller, Zhichun Li, Yan Chen, Yan Gao, Ashish Gupta, Yin Zhang,
  Peter~A. Dinda, Ming~Yang Kao, and Gokhan Memik.
\newblock {Reversible sketches: Enabling monitoring and analysis over
  high-speed data streams}.
\newblock {\em IEEE/ACM Transactions on Networking}, 15(5):1059--1072, 2007.

\bibitem{Cormode.2005}
Graham Cormode and S.~Muthukrishnan.
\newblock {What's new: Finding significant differences in network data
  streams}.
\newblock {\em IEEE/ACM Transactions on Networking}, 13(6):1219--1232, 2005.

\bibitem{Li.2016b}
Yuliang Li, Rui Miao, Changhoon Kim, and Minlan Yu.
\newblock {FlowRadar: A Better NetFlow for Data Centers}.
\newblock In {\em NSDI}, pages 311--324, 2016.

\bibitem{Huang2017}
Qun Huang, Xin Jin, Patrick P.~C. Lee, Runhui Li, Lu~Tang, Yi-Chao Chen, and
  Gong Zhang.
\newblock {SketchVisor: Robust Network Measurement for Software Packet
  Processing}.
\newblock In {\em Proceedings of the Conference of the ACM Special Interest
  Group on Data Communication}, pages 113--126, 2017.

\bibitem{Mirkovic.2015}
Curtis Yu, Cristian Lumezanu, Abhishek Sharma, Qiang Xu, Guofei Jiang, and
  Harsha~V. Madhyastha.
\newblock {Software-Defined Latency Monitoring in Data Center Networks: Passive
  and Active Measurement}.
\newblock In {\em Computer Science (including subseries Lecture Notes in
  Artificial Intelligence and Lecture Notes in Bioinformatics)}, pages
  360--372, 2015.

\bibitem{T.Mizrahi.2016}
Tal Mizrahi and Yoram Moses.
\newblock {The case for Data Plane Timestamping in SDN}.
\newblock In {\em IEEE INFOCOM}, pages 856--861, 2016.

\bibitem{He.2015}
Keqiang He, Junaid Khalid, Aaron Gember-Jacobson, Sourav Das, Chaithan Prakash,
  Aditya Akella, Li~Erran Li, and Marina Thottan.
\newblock {Measuring control plane latency in SDN-enabled switches}.
\newblock In {\em ACM SIGCOMM Symposium on Software Defined Networking
  Research}, pages 1--6, 2015.

\bibitem{Megyesi.2016}
Péter Megyesi, Alessio Botta, Giuseppe Aceto, Antonio Pescap{\`{e}}, and
  Sándor Moln{\'{a}}r.
\newblock {Available bandwidth measurement in software defined networks}.
\newblock In {\em Proceedings of the 31st Annual ACM Symposium on Applied
  Computing}, pages 651--657, 2016.

\bibitem{Popa.2013}
Lucian Popa, Praveen Yalagandula, Sujata Banerjee, Jeffrey~C. Mogul, Yoshio
  Turner, and Jose~Renato Santos.
\newblock {ElasticSwitch: Practical Work-conserving Bandwidth Guarantees for
  Cloud Computing}.
\newblock {\em ACM SIGCOMM Computer Communication Review}, 43(4):351--362,
  2013.

\bibitem{G.Aceto.2017}
Giuseppe Aceto, Valerio Persico, Antonio Pescap{\'{e}}, and Giorgio Ventre.
\newblock {SOMETIME: Software defined network-basec Available Bandwidth
  measurement in MONROE}.
\newblock In {\em Proceedings of the 1st Network Traffic Measurement and
  Analysis Conference}, 2017.

\bibitem{R.Wang.2016}
Runxin Wang, Simone Mangiante, Alan Davy, Lei Shi, and Brendan Jennings.
\newblock {QoS-aware multipathing in datacenters using effective bandwidth
  estimation and SDN}.
\newblock In {\em International Conference on Network and Service Management},
  pages 342--347, 2017.

\bibitem{F.Pakzad.2014}
Farzaneh Pakzad, Marius Portmann, Wee~Lum Tan, and Jadwiga Indulska.
\newblock {Efficient topology discovery in software defined networks}.
\newblock In {\em 8th International Conference on Signal Processing and
  Communication Systems}, 2014.

\bibitem{S.Khan.2017}
Suleman Khan, Abdullah Gani, Ainuddin~Wahid Abdul~Wahab, Mohsen Guizani, and
  Muhammad~Khurram Khan.
\newblock {Topology Discovery in Software Defined Networks: Threats, Taxonomy,
  and State-of-the-Art}.
\newblock {\em IEEE Communications Surveys and Tutorials}, 19(1):303--324,
  2017.

\bibitem{Ochoa-Aday2016}
Leonardo Ochoa-Aday, Cristina Cervell{\'{o}}-Pastor, and Adriana
  Fern{\'{a}}ndez-Fern{\'{a}}ndez.
\newblock {Discovering the Network Topology: An Efficient Approach for SDN}.
\newblock {\em Adcaij: Advances in Distributed Computing and Artificial
  Intelligence Journal}, 5(2), 2016.

\bibitem{Dai.2016}
Mian Dai, Guang Cheng, and Yuxiang Wang.
\newblock {Detecting Network Topology and Packet Trajectory with SDN-enabled
  FPGA Platform}.
\newblock In {\em Proceedings of the 11th International Conference on Future
  Internet Technologies}, volume~15, pages 7--13, 2016.

\bibitem{W.Huang.2014}
Wun~Yuan Huang, Ta~Yuan Chou, Jen~Wei Hu, and Te~Lung Liu.
\newblock {Automatical end to end topology discovery and flow viewer on SDN}.
\newblock In {\em IEEE 28th International Conference on Advanced Information
  Networking and Applications Workshops}, pages 910--915, 2014.

\bibitem{Boutaba.2018}
Raouf Boutaba, Mohammad~A. Salahuddin, Noura Limam, Sara Ayoubi, Nashid
  Shahriar, Felipe Estrada-Solano, and Oscar~M. Caicedo.
\newblock {A comprehensive survey on machine learning for networking:
  evolution, applications and research opportunities}.
\newblock {\em Journal of Internet Services and Applications}, 9(1), 2018.

\bibitem{Alsheikh2014}
M~A Alsheikh, S~Lin, D~Niyato, and H~Tan.
\newblock {Machine Learning in Wireless Sensor Networks: Algorithms,
  Strategies, and Applications}.
\newblock {\em IEEE Communications Surveys {\&} Tutorials}, 16(4):1996--2018,
  2014.

\bibitem{Bkassiny2013}
M~Bkassiny, Y~Li, and S~K Jayaweera.
\newblock {A Survey on Machine-Learning Techniques in Cognitive Radios}.
\newblock {\em IEEE Communications Surveys {\&} Tutorials}, 15(3):1136--1159,
  2013.

\bibitem{Buczak2016}
A~L Buczak and E~Guven.
\newblock {A Survey of Data Mining and Machine Learning Methods for Cyber
  Security Intrusion Detection}.
\newblock {\em IEEE Communications Surveys {\&} Tutorials}, 18(2):1153--1176,
  2016.

\bibitem{Fadlullah2017}
Z~M Fadlullah, F~Tang, B~Mao, N~Kato, O~Akashi, T~Inoue, and K~Mizutani.
\newblock {State-of-the-Art Deep Learning: Evolving Machine Intelligence Toward
  Tomorrow’s Intelligent Network Traffic Control Systems}.
\newblock {\em IEEE Communications Surveys {\&} Tutorials}, 19(4):2432--2455,
  2017.

\bibitem{Wang2018}
M~Wang, Y~Cui, X~Wang, S~Xiao, and J~Jiang.
\newblock {Machine learning for networking: Workflow, advances and
  opportunities}.
\newblock {\em IEEE Netw}, 32, 2018.

\bibitem{cyber_risk_data}
{Cyber Risk Trust Archive}.
\newblock {https://www.impactcybertrust.org}.

\bibitem{ucc_data_archive}
{UCI KDD Archive}.
\newblock {https://kdd.ics.uci.edu/}.

\bibitem{waikato_data_archive}
{WITS: Waikato Internet Traffic Storage}.
\newblock {https://wand.net.nz/wits}.

\bibitem{fraleigh_design_deployment_passive}
Chuck Fraleigh, Christophe Diot, Bryan Lyles, and Sue Moon.
\newblock {Design and deployment of a passive monitoring infrastructure}.
\newblock In {\em Evolutionary Trends of the Internet}, pages 556--575, 2001.

\bibitem{moore_internet_traffic_classification}
Andrew~W. Moore and Denis Zuev.
\newblock {Internet traffic classification using bayesian analysis techniques}.
\newblock {\em ACM SIGMETRICS Performance Evaluation Review}, 33(1), 2005.

\bibitem{yemini_netmate}
Y~Yemini and O~Wolfson.
\newblock {NETMATE: management of complex distributed networked systems}.
\newblock In {\em Proceedings of the First International Conference on Parallel
  and Distributed Information Systems}, page 173, 1991.

\bibitem{kulkarni_weka}
Eshwari~Girish Kulkarni and Raj~B. Kulkarni.
\newblock {Weka Powerful Tool in Data Mining}.
\newblock {\em IJCA Proceedings on National Seminar on Recent Trends in Data},
  RTDM 2016(2):10--15, 2016.

\bibitem{Sisiaridis2018}
Dimitrios Sisiaridis and Olivier Markowitch.
\newblock {Reducing data complexity in feature extraction and feature selection
  for big data security analytics}.
\newblock In {\em International Conference on Data Intelligence and Security,
  ICDIS}, pages 43--48, 2018.

\bibitem{Shakya2018}
V~Shakya and R~R~S Makwana.
\newblock {Feature selection based intrusion detection system using the
  combination of DBSCAN, K-Mean++ and SMO algorithms}.
\newblock In {\em International Conference on Trends in Electronics and
  Informatics, ICEI}, pages 928--932, 2018.

\bibitem{Wang2016}
Pu~Wang, Shih~Chun Lin, and Min Luo.
\newblock {A framework for QoS-aware traffic classification using
  semi-supervised machine learning in SDNs}.
\newblock In {\em IEEE International Conference on Services Computing}, pages
  760--765, 2016.

\bibitem{Zhang2013}
J~Zhang, C~Chen, Y~Xiang, W~Zhou, and Y~Xiang.
\newblock {Internet traffic classification by aggregating correlated naive
  bayes predictions}.
\newblock {\em IEEE Trans Inf Forensic Secur}, 8, 2013.

\bibitem{amor_intrusion_detection_systems}
Nahla~Ben Amor, Salem Benferhat, and Zied Elouedi.
\newblock {Naive Bayes vs decision trees in intrusion detection systems}.
\newblock In {\em ACM symposium on Applied computing}, 2004.

\bibitem{article}
James~D. Cannady.
\newblock {Artificial neural networks for misuse detection}.
\newblock {\em Proceedings of the 21st National information systems security
  conference}, 7:368--381, 1998.

\bibitem{Chebrolu2005}
Srilatha Chebrolu, Ajith Abraham, and Johnson~P. Thomas.
\newblock {Feature deduction and ensemble design of intrusion detection
  systems}.
\newblock {\em Computers and Security}, 24(4):295--307, 2005.

\bibitem{10.1007/978-3-540-45248-5_10}
Christopher Kruegel and Thomas Toth.
\newblock {Using Decision Trees to Improve Signature-Based Intrusion
  Detection}.
\newblock In {\em Recent Advances in Intrusion Detection}, pages 173--191,
  2003.

\bibitem{Subba2016}
Basant Subba, Santosh Biswas, and Sushanta Karmakar.
\newblock {A Neural Network based system for Intrusion Detection and attack
  classification}.
\newblock In {\em Twenty Second National Conference on Communication (NCC)},
  pages 1--6, 2016.

\bibitem{Pan2003}
Z.S. Pan, SC~Chen, G.B. Hu, and D.Q. Zhang.
\newblock {Hybrid neural network and C4. 5 for misuse detection}.
\newblock In {\em International Conference on Machine Learning and
  Cybernetics}, volume~4, page 2463–2467, 2003.

\bibitem{Peddabachigari2007}
Sandhya Peddabachigari, Ajith Abraham, Crina Grosan, and Johnson Thomas.
\newblock {Modeling intrusion detection system using hybrid intelligent
  systems}.
\newblock {\em Journal of Network and Computer Applications}, 30(1):114--132,
  2007.

\bibitem{Sangkatsanee2011}
Phurivit Sangkatsanee, Naruemon Wattanapongsakorn, and Chalermpol
  Charnsripinyo.
\newblock {Practical real-time intrusion detection using machine learning
  approaches}.
\newblock {\em Computer Communications}, 34(18):2227--2235, 2011.

\bibitem{Stein:2005:DTC:1167253.1167288}
Gary Stein, Bing Chen, Annie~S. Wu, and Kien~A. Hua.
\newblock {Decision tree classifier for network intrusion detection with
  GA-based feature selection}.
\newblock In {\em Proceedings of the 43rd annual southeast regional
  conference}, volume~2. ACM, 2005.

\bibitem{forster_feedback_routing_optimizing}
A~Forster and A~L Murphy.
\newblock {FROMS: Feedback Routing for Optimizing Multiple Sinks in WSN with
  Reinforcement Learning}.
\newblock In {\em 3rd International Conference on Intelligent Sensors, Sensor
  Networks and Information}, pages 371--376, 2007.

\bibitem{hu_qelar_routing_protocol}
T~Hu and Y~Fei.
\newblock {Qelar: a machine-learning-based adaptive routing protocol for
  energy-efficient and lifetime-extended underwater sensor networks}.
\newblock {\em IEEE Trans Mob Comput}, 9(6):796--809, 2010.

\bibitem{lin_qos_aware_routing}
S~Lin, I~F Akyildiz, P~Wang, and M~Luo.
\newblock {QoS-Aware Adaptive Routing in Multi-layer Hierarchical Software
  Defined Networks: A Reinforcement Learning Approach}.
\newblock In {\em IEEE International Conference on Services Computing (SCC)},
  pages 25--33, 2016.

\bibitem{piamrat_qoe_aware_admission}
K~Piamrat, A~Ksentini, C~Viho, and J~Bonnin.
\newblock {QoE-Aware Admission Control for Multimedia Applications in IEEE
  802.11 Wireless Networks}.
\newblock In {\em IEEE 68th Vehicular Technology Conference}, pages 1--5, 2008.

\bibitem{baldo_call_admission_control}
N~Baldo, P~Dini, and J~Nin-Guerrero.
\newblock {User-driven Call Admission Control for VoIP over WLAN with a Neural
  Network based cognitive engine}.
\newblock In {\em 2nd International Workshop on Cognitive Information
  Processing}, pages 52--56, 2010.

\bibitem{Fonseca2005}
N.~Fonseca and M.~Crovella.
\newblock {Bayesian packet loss detection for TCP}.
\newblock In {\em Proceedings IEEE 24th Annual Joint Conference of the IEEE
  Computer and Communications Societies.}, volume~3, pages 1826--1837, 2005.

\bibitem{Haffner2005}
Patrick Haffner, Subhabrata Sen, Oliver Spatscheck, and D~Wang. Acas.
\newblock {Automated construction of application signatures}, 2005.

\bibitem{Ma:2006:UMP:1177080.1177123}
Justin Ma, Kirill Levchenko, Christian Kreibich, Stefan Savage, and Geoffrey~M.
  Voelker.
\newblock {Unexpected means of protocol inference}.
\newblock In {\em ACM SIGCOMM on Internet measurement}, 2006.

\bibitem{Schatzmann:2010:DHF:1879141.1879184}
Dominik Schatzmann, Wolfgang M{\"{u}}hlbauer, Thrasyvoulos Spyropoulos, and
  Xenofontas Dimitropoulos.
\newblock {Digging into HTTPS: Flow-Based classification of webmail traffic}.
\newblock In {\em Proceedings of the 10th annual conference on Internet
  measurement}, pages 322--326. ACM, 2010.

\bibitem{Jiang2007}
Hongbo Jiang, Andrew~W. Moore, Zihui Ge, Shudong Jin, and Jia Wang.
\newblock {Lightweight application classification for network management}.
\newblock In {\em Proceedings of the SIGCOMM workshop on Internet network
  management}, 2007.

\bibitem{Zhang2015}
Jun Zhang, Xiao Chen, Yang Xiang, Wanlei Zhou, and Jie Wu.
\newblock {Robust Network Traffic Classification}.
\newblock {\em IEEE/ACM Transactions on Networking}, 23(4):1257--1270, 2015.

\bibitem{Sun2007}
Runyuan Sun, Lizhi Peng, Zhenxiang Chen, Lei Zhang, and Shan Jing.
\newblock {Traffic classification using probabilistic neural networks}.
\newblock In {\em International Conference on Natural Computation}, volume~4,
  pages 299--304, 2007.

\bibitem{Nguyen2012}
Thuy T.~T. Nguyen, Grenville Armitage, Philip Branch, and Sebastian Zander.
\newblock {Timely and Continuous Machine-Learning-Based Classification for
  Interactive IP Traffic}.
\newblock {\em IEEE/ACM Transactions on Networking}, 20(6):1880--1894, 2012.

\bibitem{dainotti_early_classification_traffic}
Alberto Dainotti, Antonio Pescap, and Carlo Sansone.
\newblock {Early Classification of Network Traffic through
  Multi-classification}.
\newblock In {\em Traffic Monitoring and Analysis}, pages 122--135, 2011.

\bibitem{Donato2014}
Walter Donato, Antonio Pescap{\'{e}}, and Alberto Dainotti.
\newblock {Traffic identification engine: An open platform for traffic
  classification}.
\newblock {\em IEEE Network}, 28(2):56--64, 2014.

\bibitem{liu_svm_based_classification}
C~Liu, Y~Chang, C~Tseng, Y~Yang, M~Lai, and L~Chou.
\newblock {SVM-based Classification Mechanism and Its Application in SDN
  Networks}.
\newblock In {\em International Conference on Communication Software and
  Networks (ICCSN)}, pages 45--49, 2018.

\bibitem{leng_flow_table_compression}
B~Leng, L~Huang, Chunming Qiao, and H~Xu.
\newblock {A decision-tree-based on-line flow table compressing method in
  Software Defined Networks}.
\newblock In {\em IEEE/ACM 24th International Symposium on Quality of Service
  (IWQoS)}, pages 1--2, 2016.

\bibitem{kolomvatsos_uncertainty_driven_ensemble}
K~Kolomvatsos, C~Anagnostopoulos, A~K Marnerides, Q~Ni, S~Hadjiefthymiades, and
  D~P Pezaros.
\newblock {Uncertainty-driven ensemble forecasting of QoS in Software Defined
  Networks}.
\newblock In {\em 2017 IEEE Symposium on Computers and Communications (ISCC)},
  pages 1284--1289, 2017.

\bibitem{tang_neural_network_intrusion}
T~A Tang, L~Mhamdi, D~McLernon, S~A~R Zaidi, and M~Ghogho.
\newblock {Deep Recurrent Neural Network for Intrusion Detection in SDN-based
  Networks}.
\newblock In {\em IEEE Conference on Network Softwarization and Workshops
  (NetSoft)}, pages 202--206, 2018.

\bibitem{al2020arabic_UOB3}
Abdulhakeem~Qusay Al-Bayati, Ahmed~S Al-Araji, and Saman~Hameed Ameen.
\newblock Arabic sentiment analysis (asa) using deep learning approach.
\newblock {\em Journal of Engineering}, 26(6):85--93, 2020.

\bibitem{mao_network_traffic_control}
Bomin Mao, Fengxiao Tang, Zubair~Md Fadlullah, Nei Kato, Osamu Akashi, Takeru
  Inoue, and Kimihiro Mizutani.
\newblock {A Novel Non-Supervised Deep-Learning-Based Network Traffic Control
  Method for Software Defined Wireless Networks}.
\newblock {\em IEEE Wireless Communications}, 25(4):74--81, 2018.

\bibitem{chavula_sdn_reinforcement_learning}
J~Chavula, M~Densmore, and H~Suleman.
\newblock {Using SDN and reinforcement learning for traffic engineering in
  UbuntuNet Alliance}.
\newblock In {\em International Conference on Advances in Computing and
  Communication Engineering (ICACCE)}, pages 349--355, 2016.

\bibitem{liu_reinforcement_ddos_flooding}
Y~Liu, M~Dong, K~Ota, J~Li, and J~Wu.
\newblock {Deep Reinforcement Learning based Smart Mitigation of DDoS Flooding
  in Software-Defined Networks}.
\newblock In {\em IEEE 23rd International Workshop on Computer Aided Modeling
  and Design of Communication Links and Networks (CAMAD)}, pages 1--6, 2018.

\bibitem{zhang_qplacement_reinforcement_learning}
Z~Zhang, L~Ma, K~K Leung, L~Tassiulas, and J~Tucker.
\newblock {Q-Placement: Reinforcement-Learning-Based Service Placement in
  Software-Defined Networks}.
\newblock In {\em IEEE 38th International Conference on Distributed Computing
  Systems (ICDCS)}, pages 1527--1532, 2018.

\bibitem{kim_congestion_prevention_mechanism}
S~Kim, J~Son, A~Talukder, and C~S Hong.
\newblock {Congestion prevention mechanism based on Q-leaning for efficient
  routing in SDN}.
\newblock In {\em International Conference on Information Networking (ICOIN)},
  pages 124--128, 2016.

\bibitem{min_dynamic_switch_migration}
Z~Min, Q~Hua, and Z~Jihong.
\newblock {Dynamic switch migration algorithm with Q-learning towards scalable
  SDN control plane}.
\newblock In {\em 9th International Conference on Wireless Communications and
  Signal Processing (WCSP)}, pages 1--4, 2017.

\bibitem{qiu_blockchain_deep_qlearning}
C~Qiu, F~R Yu, H~Yao, C~Jiang, F~Xu, and C~Zhao.
\newblock {Blockchain-Based Software-Defined Industrial Internet of Things: A
  Dueling Deep Q-Learning Approach}.
\newblock {\em IEEE Internet of Things Journal}, 2018.

\bibitem{zhang_resource_saving_replication}
L~Zhang, Y~Wang, X~Zhong, W~Li, and S~Guo.
\newblock {Resource-saving replication for controllers in multi controller SDN
  against network failures}.
\newblock In {\em IEEE/IFIP Network Operations and Management Symposium}, pages
  1--7, 2018.

\bibitem{adami_novel_sdn_controller}
D~Adami, S~Giordano, M~Pagano, and G~Portaluri.
\newblock {A novel SDN controller for traffic recovery and load balancing in
  data centers}.
\newblock In {\em IEEE 21st International Workshop on Computer Aided Modelling
  and Design of Communication Links and Networks (CAMAD)}, pages 77--82, 2016.

\bibitem{shu_traffic_measurement_management}
Z~Shu, J~Wan, J~Lin, S~Wang, D~Li, S~Rho, and C~Yang.
\newblock {Traffic engineering in software-defined networking: Measurement and
  management}.
\newblock {\em IEEE Access}, 4:3246--3256, 2016.

\bibitem{T_SDN_load}
Tariq~Emad Ali, Ameer~Hussein Morad, and Mohammed~A Abdala.
\newblock Load balance in data center sdn networks.
\newblock {\em International Journal of Electrical and Computer Engineering
  (IJECE)}, 8(5):3086--3092, 2018.

\bibitem{hamed_novel_resource_utilization}
M~I Hamed, B~M ElHalawany, M~M Fouda, and A~S~T Eldien.
\newblock {A novel approach for resource utilization and management in SDN}.
\newblock In {\em International Computer Engineering Conference (ICENCO)},
  pages 337--342, 2017.

\bibitem{li_estimating_sdn_traffic}
D~Li, N~Dai, F~Li, C~Xing, and F~Dai.
\newblock {Estimating SDN Traffic Matrix Based on Online Informative Flow
  Measurement Method}.
\newblock In {\em International Conference on Advanced Cloud and Big Data
  (CBD)}, pages 75--80, 2017.

\bibitem{tse_sdn_enabled_core}
Simon Tse and Gagan Choudhury.
\newblock {Real-Time Traffic Management in AT{\&}amp;T's SDN-Enabled Core
  IP/Optical Network}.
\newblock In {\em Optical Fiber Communications Conference and Exposition
  (OFC)}, pages 1--3, 2018.

\bibitem{zhou_real_time_sdn}
H~Zhou, C~Wu, C~Yang, P~Wang, Q~Yang, Z~Lu, and Q~Cheng.
\newblock {SDN-RDCD: A Real-Time and Reliable Method for Detecting Compromised
  SDN Devices}.
\newblock {\em IEEE/ACM Transactions on Networking}, 26(5):2048--2061, 2018.

\bibitem{T_SDN_MGM}
Tariq Ali, Ameer Morad, and Mohammed Abdala.
\newblock Traffic management inside software-defined data centre networking.
\newblock {\em Bulletin of Electrical Engineering and Informatics},
  9(5):2045--2054, 2020.

\bibitem{su_costa_cross_layer}
Z~Su, T~Wang, and M~Hamd.
\newblock {COSTA: Cross-layer optimization for sketch-based software defined
  measurement task assignment}.
\newblock In {\em IEEE 23rd International Symposium on Quality of Service
  (IWQoS)}, pages 183--188, 2015.

\bibitem{kaplan_development_sketch_based}
L~Kaplan and T~Halagan.
\newblock {Development sketch-based tool for creation and scaling of
  virtualized SDN infrastructure}.
\newblock In {\em International Conference on Emerging eLearning Technologies
  and Applications (ICETA)}, pages 1--5, 2015.

\bibitem{monshizadeh_adaptive_detection_prevention}
M~Monshizadeh, V~Khatri, and R~Kantola.
\newblock {An adaptive detection and prevention architecture for unsafe traffic
  in SDN enabled mobile networks}.
\newblock In {\em IFIP/IEEE Symposium on Integrated Network and Service
  Management (IM)}, pages 883--884, 2017.

\bibitem{rebecchi_traffic_monitoring_ddos}
F~Rebecchi, J~Boite, P~Nardin, M~Bouet, and V~Conan.
\newblock {Traffic monitoring and DDoS detection using stateful SDN}.
\newblock In {\em IEEE Conference on Network Softwarization (NetSoft)}, pages
  1--2, 2017.

\end{thebibliography}

\end{document}